\documentclass[default, twocolumn]{aastex701}
\usepackage{comment}
\usepackage[T1]{fontenc}
\usepackage{amsmath}

\newcommand\aastex{AAS\TeX}

\begin{document}

\title{Carbonaceous Chondrites provide evidence for late-stage planetesimal formation in a pressure bump}

\author[0009-0008-3256-9564]{Nerea~Gurrutxaga}
\affiliation{Max Planck Institute for Solar System Research, Justus-von-Liebig-Weg 3, Göttingen, 37077, Germany}
\email{gurrutxaga@mps.mpg.de}  

\author[0000-0002-9128-0305]{Joanna~Dr{\k{a}}{\.z}kowska} 
\affiliation{Max Planck Institute for Solar System Research, Justus-von-Liebig-Weg 3, Göttingen, 37077, Germany}
\email{fakeemail2@google.com}

\author[0000-0002-2451-9574]{Vignesh~Vaikundaraman}
\affiliation{Max Planck Institute for Solar System Research, Justus-von-Liebig-Weg 3, Göttingen, 37077, Germany}
\email{fakeemail3@google.com}

\author[0000-0003-4657-5961]{Thorsten~Kleine}
\affiliation{Max Planck Institute for Solar System Research, Justus-von-Liebig-Weg 3, Göttingen, 37077, Germany}
\email{fakeemail4@google.com}


\begin{abstract}

Carbonaceous chondrites are samples from planetesimals that formed 2–4 million years after solar system formation began. They consist of distinct dust components formed at different times and locations in the accretion disk and whose abundances in carbonaceous chondrites vary over planetesimal formation time. The mechanism that led to this time-varied accretion is not understood, but is critical for understanding late-stage planetesimal formation. Using a two-dimensional Monte Carlo simulation of dust evolution, we show that differences in dust filtering and delivery rates of distinct dust components to a planet-induced pressure bump in the disk reproduce the observed compositions and formation ages of the carbonaceous chondrites. This implies that carbonaceous chondrites likely formed in a single, long-lived dust trap, most likely outside of Jupiter’s orbit. Because differentiated meteorites, which sample an earlier generation of planetesimals, exhibit similar isotopic variability as the chondrites, they likely have also formed in dust traps, implying these structures were the dominant site for planetesimal formation in the solar system.

\end{abstract}

\keywords{\uat{Meteorite composition}{1037} --- \uat{Carbonaceous chondrites}{200} --- \uat{Small Solar System bodies}{1469} --- \uat{Protoplanetary disks}{1300} --- \uat{Planet formation}{1241} --- \uat{Monte Carlo methods}{2238}}


\section{Introduction} 
The initial stages of planet formation span an enormous range of masses and involve the growth of tiny dust grains into km-sized precursor objects called planetesimals \citep[e.g.,][]{Johansen2014}. As fragments of leftover planetesimals, meteorites provide a unique record of the conditions and timescales of the initial stages of planet formation in the solar system. Their ages and compositions reveal that planetesimal formation occurred at distinct locations and throughout almost the entire lifetime of the solar accretion disk \citep[e.g.,][]{Kleine2020}. Chondrite meteorites derive from parent bodies that formed late, between 2–4 million years (Myr) after solar system formation began \citep[e.g., ][]{Sugiura2014, Fukuda2022}, and did not undergo significant melting and chemical differentiation. As such, they preserve a direct record of the distinct dust components that were combined to form their parent planetesimals. Chondrites, therefore, allow studying the processes by which small dust grains have been accreted into the first sizable planetary objects.

Among the chondrites, carbonaceous chondrites are the most primitive and diverse objects. They consist of refractory inclusions and chondrules embedded into a fine-grained matrix \citep{Scott2014}. Refractory inclusions such as Ca-Al-rich inclusions (CAIs) are the oldest dated solids of the solar system and, as high-temperature condensates, they are thought to have formed near the young Sun before being transported outward by disk spreading or outflows \citep{Cuzzi2003, Ciesla2007}. Chondrules are sub-millimeter-sized igneous spherules that formed by transient heating events of dust aggregates 2-4 Myr after CAI formation \citep[e.g.,][]{Krot2009}. The matrix contains relatively unprocessed submicron-sized mineral grains, and is compositionally similar to the chemically most primitive chondrites, the Ivuna-type (CI) carbonaceous chondrites. Variations in the relative abundance of these three constituent components define the different groups of carbonaceous chondrites, each representing planetesimals that formed at different times and have distinct compositions \cite[e.g.,][]{Alexander2019, Hellmann2020}. Moreover, refractory inclusion and chondrule abundances are correlated within each chondrite group, and matrix-poor chondrites appear to have formed earlier than matrix-rich chondrites \citep{Hellmann2023}. Only the Renazzo-type (CR) chondrites deviate from this trend because they are rich in chondrules but poor in refractory inclusions and matrix despite their relatively late formation time \citep{Schrader2017, Budde2018}.

Theoretical models indicate that planetesimal formation requires high concentrations of solid particles relative to gas \citep{Youdin2005, Lim2024}. This condition is difficult to achieve in disks with ages of $2-4\,\mathrm{Myr}$, when the carbonaceous chondrites formed, because dust is efficiently depleted on a characteristic timescale of less than one Myr \citep{Drazkowska2023}. A common solution to this problem is to invoke dust enrichment and planetesimal formation in pressure bumps that, especially during the later stages of disk evolution, form in response to the presence of giant planets \citep{Pinilla2012, Stammler2019, Eriksson2020}. Indeed, it has been suggested that carbonaceous chondrites formed in a pressure bump outside Jupiter’s orbit \citep{Desch18}, and that their formation in a single pressure bump could potentially explain the trapping of refractory inclusions and chondrules (or their precursors) over a prolonged time \citep{Hellmann2023}. However, whether the temporal variations in the abundances of refractory inclusions and chondrules over matrix across the carbonaceous chondrite groups can be produced in a pressure bump has not yet been quantitatively explored. Addressing this question is of considerable interest, as it would allow determining the mechanism by which distinct dust components have been accreted into planetesimals, and how this mechanism may have changed over time.

\begin{figure*}
\centering
\includegraphics[width=0.9\textwidth]{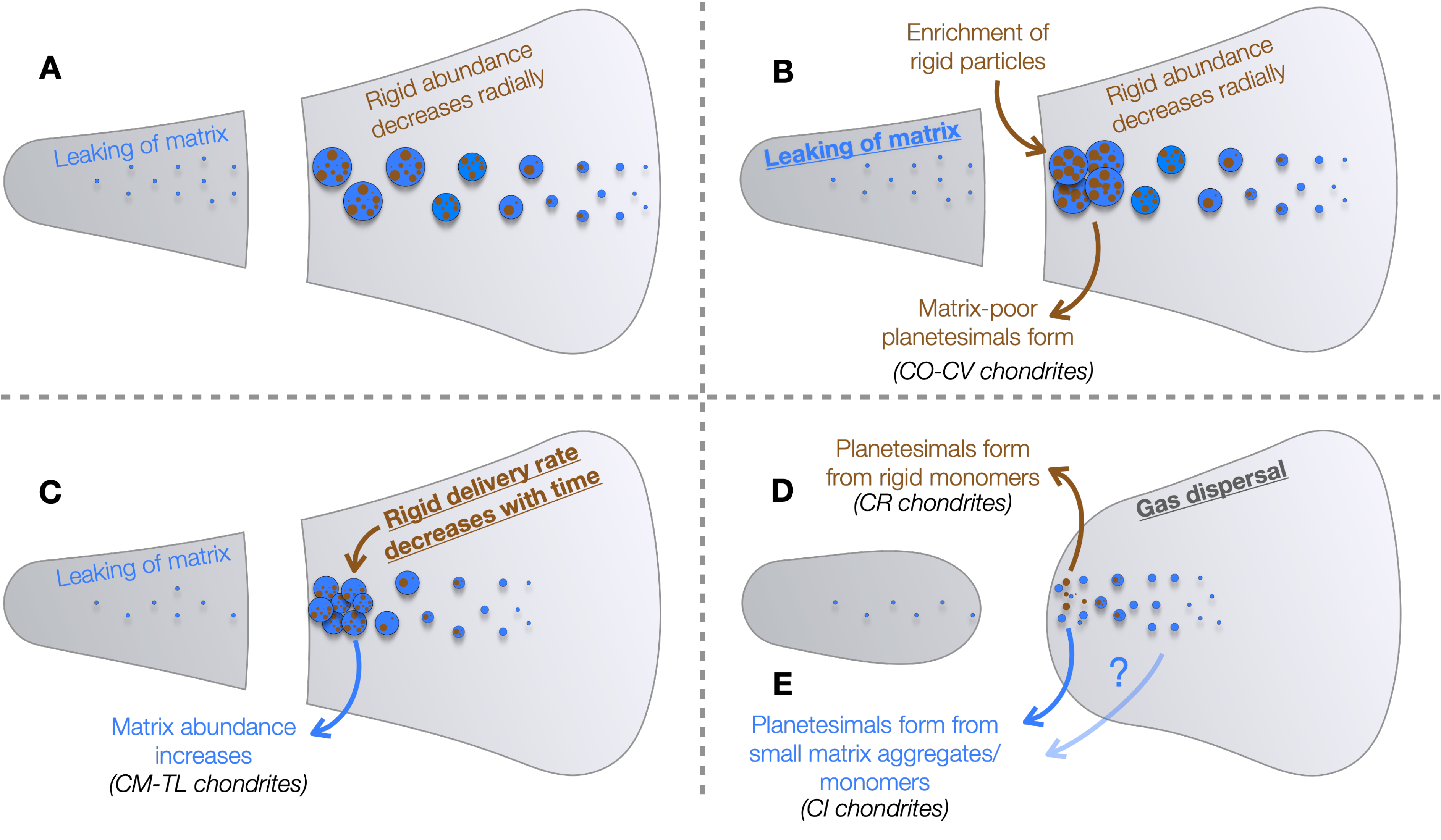}
\caption{Schematic of the model for carbonaceous chondrite formation. We assume that chondrules and refractory inclusions are rigid particles, and that the matrix is fragile. Rigid and fragile materials can stick together to form larger pebbles. We highlight the dominant process at different times in bold and underlined. (\textbf{A}) Initially, $\sim$2 Myr after CAI formation, a gap is opened by a Jupiter-like planet. (\textbf{B}) Mostly fragile material passes through the gap, enriching the pressure bump with rigid particles. Planetesimal formation begins once the pebble-to-gas ratio becomes sufficiently high. (\textbf{C}) Then, the delivery of rigid particles to the pressure bump declines over time due to the faster radial drift (and thereby faster depletion in the outer disk) of the largest rigid monomers. (\textbf{D}) Afterwards, photoevaporation reduces the gas surface density by orders of magnitude and sub-millimeter-sized rigid particles reach pebble Stokes numbers. (\textbf{E}) Finally, as the gas density continues to decrease, particle sizes are further reduced to micrometer sizes. At this late stage, planetesimal formation may extend across the disk as the gap expands during photoevaporation.}\label{fig:model}
\end{figure*}

Here we present a model that combines gas disk evolution, planetesimal formation, and dust evolution, taking into account collisions between the different dust components present in carbonaceous chondrites. Figure\,\ref{fig:model} summarizes our model. In Section~\ref{sec:methods}, we describe the gas disk model, the dust coagulation and fragmentation model for multiple dust components, and the setup of the simulations. In Section~\ref{sec:results}, we present the results of our simulations. In Section~\ref{sec:discussion}, we compare our results to the observed properties of carbonaceous chondrites, and we discuss the implications for understanding the origin of other chondrite parent bodies, as well as the origin of other meteorite parent bodies that formed earlier than the chondrites. We summarize our findings in Section \ref{sec:conclusion}.

\section{Methods}\label{sec:methods}

\subsection{Disk model}

\begin{figure}[h]
\centering
\includegraphics[width=0.4\textwidth]{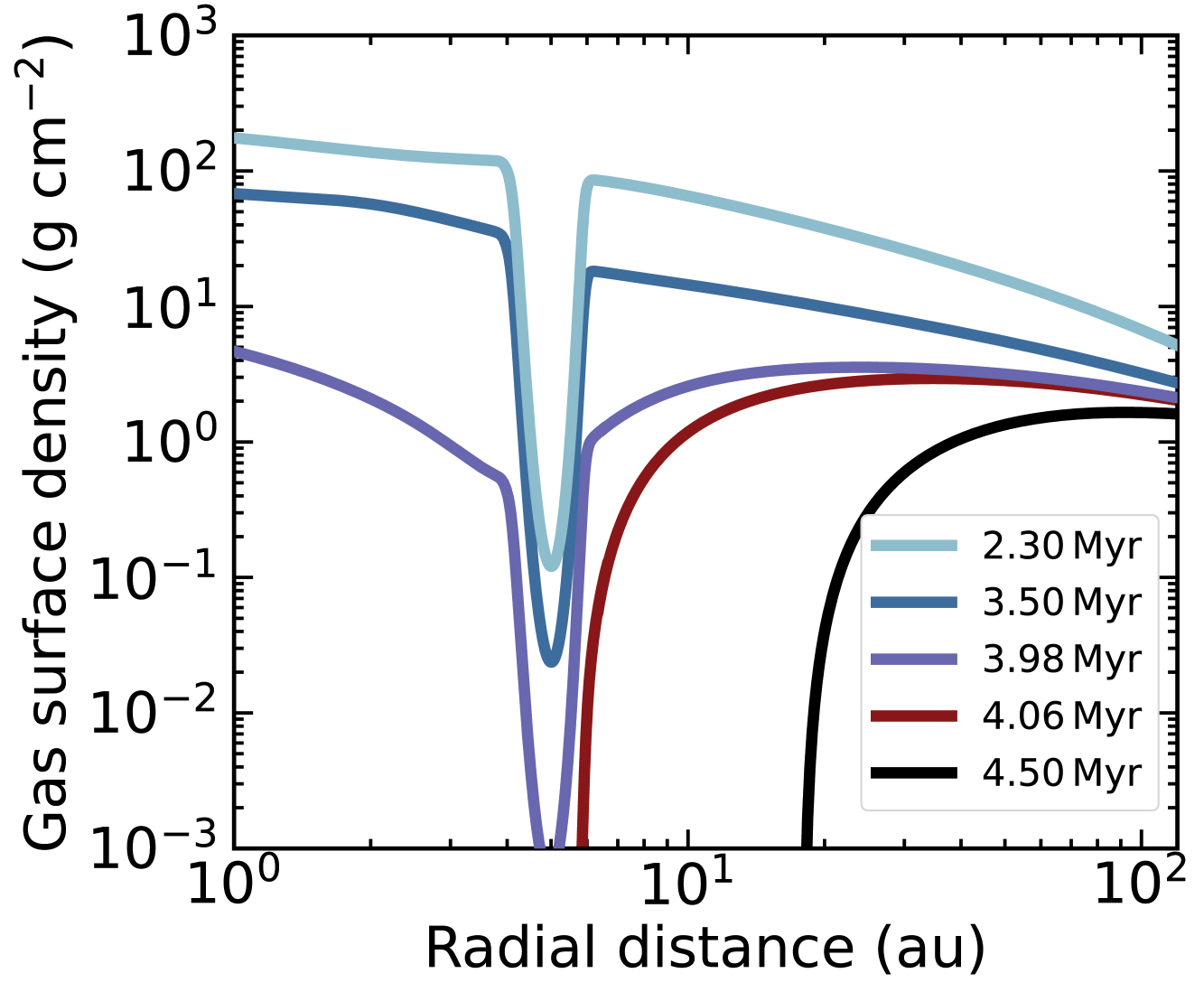}
\caption{Gas surface density across the disk at different stages. A Jupiter-like planet opens a gap at $5\,\mathrm{AU}$. Disk dispersal by internal photoevaporation is included. Times are given relative to CAI formation, and the time difference between disk formation and CAI formation in our model is around $0.19\,\mathrm{Myr}$.}\label{fig:gas}
\end{figure}

We build a one-dimensional gas disk model that includes both disk formation and a planetary gap (Figure\,\ref{fig:gas}). We solve the viscous disk equations incorporating a source function that describes disk formation from the collapse of a rotating molecular cloud core using the publicly available code \texttt{DD-Diskevol}\footnote{ \url{https://github.com/astrojoanna/DD-diskevol}} \citep{Hueso2005,Drazkowska2018}. Furthermore, we consider gas removal in the late evolutionary stages driven by internal photoevaporation, using a mass-loss rate derived by \cite{Picogna2021}. The evolution of the gas surface density is given by:
\begin{equation}
    \frac{\partial \Sigma_{\rm{g}}}{\partial t} = \frac{3}{r}\frac{\partial}{\partial r}\left[ r^{1/2}\frac{\partial}{\partial r} \left( \nu \Sigma_{\rm{g}}r^{1/2}\right)\right] + \dot{\Sigma}_{\rm{g, inf}} - \dot{\Sigma}_{\rm{g, PE}}\,,
\end{equation}
where $r$ is the radial distance from the central star, $\Sigma_{\rm{g}}$ is the gas surface density, $\nu$ is the viscosity, $\dot{\Sigma}_{\rm{g, inf}}$ is the infall rate \citep[Eq.~6 in][]{Hueso2005}, and $\dot{\Sigma}_{\rm{g, PE}}$ is the mass-loss rate due to internal photoevaporation \citep[Eqs.~6 and 7 in][]{Picogna2021}. We introduce the mass-loss rate term at $1\,\mathrm{Myr}$ after the onset of disk formation to avoid interfering with the early disk buildup phase, which starts with a gas surface density set to a floor value. As the disk evolves, the star mass increases in our model \citep[see Fig.~1 in][]{Drazkowska2018}. However, the photoevaporation prescription of \citet{Picogna2021} is given for a fixed star mass. Since photoevaporation becomes significant at late evolutionary stages, when the star has already acquired most of its final mass, we adopt the photoevaporation model for a solar-mass star \citep[see Table 2 in][]{Picogna2021}.

We assume that the disk is heated by viscous dissipation and stellar irradiation and we adopt the opacities from \cite{Bell1994} \citep[for further details, see][ their Section 3.4.2.]{Hueso2005}. For the vertical structure, we assume a Gaussian gas density profile and isothermal temperature. To describe the viscosity $\nu$ we employ the $\alpha$-disk model \citep{Shakura1973} given by:
\begin{equation}
    \nu = \alpha_{\rm{acc}} c_{\rm{s}} H \,,
\end{equation}
where $\alpha_{\rm{acc}} $ is the dimensionless accretion rate, $c_{\rm{s}}$ is the sound speed, and $H$ is the vertical scale-height. $\alpha_{\rm{acc}} $$\sim$$\, 10^{-2}-10^{-3}$ explains best the observed gas accretion rates toward young Sun-like stars and disk lifetimes of a few Myr \citep{Hartmann1998}. We choose $\alpha_{\rm{acc}}$$\,=$$\,5\times 10^{-3}$, since this value corresponds to a disk lifetime of $4-5\,\mathrm{Myr}$ after CAI formation, consistent with estimates of the solar accretion disk lifetime derived from meteorite paleomagnetic studies \citep[][]{Weiss2021}. Observations from dust settling, however, suggest that turbulence in the midplane must be lower ($\alpha_{\rm{t}} \lesssim  10^{-4}$) \citep{Pinte2023}. Therefore, we distinguish between the turbulence parameter $\alpha_{\rm{t}}$$\,=$$\,10^{-4}$ and the global accretion parameter $\alpha_{\rm{acc}}$$\,=$$\,5\times 10^{-3}$.

To estimate the time difference between the onset of infall and the formation time of CAIs, we assume that CAIs condense at temperatures between $1400$ and $1800\,\mathrm{K}$. We then calculate the time at which the disk reaches this temperature range at the centrifugal radius \citep[][their Equation 5]{Hueso2005}, as material can be transported outward beyond this location \citep{Jongejan2023}. For our chosen parameters, the time difference between the onset of infall and the formation of the first CAIs is $0.19\,\mathrm{Myr}$.

Globally the gas disk rotates at sub-Keplerian velocities, since the radially decreasing pressure gradient exerts an outward force that partially counteracts the gravitational force of the central star. The deviation from the Keplerian rotation is calculated by:
\begin{equation}\label{eq:deltav}
    \Delta v = -\frac{1}{2} \frac{H}{r} \frac{\partial \ln P}{\partial \ln r} c_s,
\end{equation}
where $ P $ is the gas pressure. Due to the relative velocity $\Delta v$ between the orbital motion of the gas and that of solid particles, the gas exerts a drag force on the solids. Small dust particles respond rapidly to this drag force and remain well coupled to the gas motion. In contrast, larger solid particles, referred to as pebbles, do not respond quickly enough to the drag force and therefore undergo inward radial drift. Pressure bumps or substructures in the gas disk, such as planetary gaps, can alter the radial pressure profile, creating local pressure maxima that halt the inward drift of pebbles.

Our model includes a gap opened by a Jupiter-like planet in a fixed orbit at $5 \,\mathrm{AU}$, a scenario that has been previously studied \citep[e.g.,][]{Desch18, Weber2018, Haugbolle2019, Jongejan2023, Stammler2023, Clepper2025, Pfeil2025}. In reality, the gap may result from overlapping gaps formed by Jupiter and Saturn in resonance \citep{Morbidelli2007}, or by a migrating Jupiter alone. Nevertheless, it is expected that the overall dust dynamics remain qualitatively consistent across these scenarios \citep[see discussion in ][]{Weber2018}. To mimic the presence of a Jupiter-like planet, we adopt the model from \cite{Desch18} and \cite{Jongejan2023}. We increase the $\alpha_{\rm{acc}}$ value such as
\begin{equation}
    \alpha^{\prime} = \alpha_{\rm{acc}} + (\alpha_{\rm{peak}}-\alpha_{\rm{acc}})e^{-x^{2}}\,,
\end{equation}
where $x$$\,=$$\,(r-r_{\rm{J}})/R_{\rm{H}}$, with $R_{\rm{H}}$$\,=$$\,r_{\rm{J}}(M_{\rm{J}}/3M_{\odot})^{1/3}$ being the Hill radius of Jupiter. We start smoothly opening a gap from $0.6\,\mathrm{Myr}$ once the infall terminates, and linearly increase the value of $\alpha_{\rm{peak}}$ to $1000\,\alpha_{\rm{acc}}$ until $1.5\,\mathrm{Myr}$. We use a simpler approach than \cite{Desch18} and \cite{Jongejan2023}, as our simulations of dust evolution begin $2\,\mathrm{Myr}$ after CAI formation and assume that Jupiter had already formed (see Section~\ref{sec:dust0}). Saturn is not included, although it is expected to have formed before disk dispersal. Previous studies showed that incorporating an additional Saturn-like planet does not necessarily inhibit the delivery of material to the outer region of Jupiter's gap \citep{Jongejan2023}. This depends on Saturn's formation history, and here we assume that Saturn did not perturb its vicinity significantly to affect dust evolution. All parameters employed to calculate gas evolution are listed in Table~\ref{tab:gas}.

\subsection{Dust evolution model}\label{sec:dust0}

Some prior studies have investigated dust trapping
with fixed particle sizes outside a planetary gap \citep{Desch18, Jongejan2023}, distinguishing
between large refractory inclusions and micrometer-sized matrix. These studies show that refractory inclusions are efficiently trapped in the pressure bump and that matrix crosses the gap, but in reality, different dust components are expected to collide and stick together, forming larger pebble-sized aggregates \citep{Beitz2012, Machii2013, Gunkelmann2017, Umstatter2021}. If micrometer-sized matrix monomers grow to larger aggregates, the matrix comprises both small fragments that cross
the gap and pebbles that remain trapped. Moreover, dust growth is typically necessary to reach pebble sizes large enough to aerodynamically decouple from the gas and initiate planetesimal formation \citep{Bai2010, Lim2024}. Other studies have modeled dust evolution using single-component dust growth and fragmentation \citep{Stammler2023, Pfeil2025}, showing that while pebbles are trapped, fragmentation allows small dust to leak through the gap via advection and diffusion. These single-component models however are not directly applicable to the study of carbonaceous chondrite formation, because chondrules and refractory inclusions are coarser, more crystalline, and more resistant to fragmentation than the matrix \citep{Scott2014}, and consequently, dust cannot be treated as a single component.

To fully account for these effects, we model collisions between different dust components and the more fragile nature of the matrix compared to refractory inclusions and chondrules. \cite{Hellmann2023} found that the abundance of refractory inclusions and chondrules in each carbonaceous chondrite group is correlated (except in CR chondrites, see Section~\ref{sec:discussion}), and suggests that the correlation may have been established before chondrule formation. Since our goal is to study the temporal variations in the abundances of refractory inclusions and chondrules over matrix, for simplicity, we treat refractory inclusions and chondrules equally. We distinguish between two types of materials, fragile and rigid, which correspond to matrix and refractory inclusions/chondrules, respectively.

We model dust evolution using a two-dimensional Monte Carlo code that accounts for both dust transport and collisions \citep{Zsom2008, Drazkowska2013, Vaikundaraman2025}. We compute the evolution of $N_{r}$ representative particles, each representing $N_{i}$ identical physical particles. We track the properties of each representative particle, including the particle mass $m_i$, spatial position $(r_i, z_i)$, and composition (see the next section). Following \cite{Zsom2008}, each representative particle corresponds to a group of physical particles (or a "swarm") of constant total mass $M_{\mathrm{swarm}}$. Consequently, a representative particle $i$ with particle mass $m_i$ represents $N_{i}\,$$=\,$$M_{swarm}/m_{i}$ physical particles. Each representative particle is subject to radial drift, advection, vertical settling, and diffusion, following the prescriptions in \cite{Drazkowska2013}.

To calculate collisions between nearby particles, we build an adaptive grid every timestep \citep{Drazkowska2013}. The grid is divided into $n_{r}$ radial and $n_{z}$ vertical bins, resulting in a total of $n_{r} \times n_{z}$ cells. Each cell contains an equal number of representative particles. Collisions are computed only between representative particles within the same cell. For each representative particle $i$, the collision rate with a physical particle represented by particle $j$ is given by
\begin{equation}
C_{i,j} = \frac{N_{j},\Delta v_{i,j}\sigma_{i,j}}{V}\,,
\end{equation}
where $N_{j}$ is the number of physical particles represented by particle $j$, $\Delta v_{i,j}$ is the relative velocity between the particles, $\sigma_{i,j}$ is their geometric collision cross-section, and $V$ is the volume of the cell. Following a Monte Carlo approach, the time interval between collision events and the colliding particle pairs are determined by random sampling \citep{Gillespie1975}. In each collision, a representative particle $i$ interacts with a physical particle represented by $j$, and only particle $i$ is updated after the interaction \citep{Zsom2008}. Depending on their relative velocity $\Delta v_{i,j}$, the collision results in coagulation, fragmentation, or bouncing (see the next section).

When the particle size distribution is broad, collisions between particles $i$ and $j$ with very different masses may occur frequently, but a single collision changes the mass of the larger particle insignificantly. To speed up the Monte Carlo algorithm, we can approximate multiple physical collisions as a single effective collision. \cite{Zsom2008} introduced a parameter, $dm_{\rm{max}}$, to group collisions between particles of very different masses. If the mass-ratio $m_{j}/m_{i}$ is lower than $dm_{\rm{max}}$, the collision rate $C_{i,j}$ is then modified as follows:
\begin{equation}
    C^{\prime}_{i,j} = \frac{m_{j}/m_{i}}{dm_{\rm{max}}}C_{i,j}\,,
\end{equation}
and if collision happens, $\frac{m_{i}}{m_{j}}dm_{\rm{max}}$ particles of $j$ collide with particle $i$. We set $dm_{\mathrm{max}}=0.01$ as a fiducial value, which in local simulations introduces a mass error of $1\%$. However, in global disks, grouping collisions may be unrealistic if particles move across grid cells on timescales shorter than those required to complete the grouped collisions. To address this, we first estimate the maximum mass-ratio that allows for grouping in global disks by equating the dust transport timescale with the grouped collision timescale, which results in:
\begin{equation}
dm_{\rm{max}, \,r}=\frac{\Delta r}{v_{r, i}}\frac{m_{j}}{m_{i}}C_{i,j}\,,
\end{equation}
where $\Delta r$ is the radial width of the grid cell. Here, the transport velocity $v_{r, i}$ accounts for all radial transport mechanisms except turbulence, which is excluded due to its stochastic nature. For the vertical transport, we replace the radial coordinate $r$ by the vertical coordinate $z$. We define an adaptive grouping parameter, $dm^{\prime}_{\mathrm{max}}$, which is the minimum between our fiducial value $dm_{\mathrm{max}}$ and the mass-ratio allowed for grouping in the global disk
\begin{equation}
dm^{\prime}_{\rm{max}}=\min\left(0.01,\; dm_{\rm{max, r}}, \;dm_{\rm{max, z}} \right)\,.
\end{equation}
This readjustment is especially important for studying dust evolution in a pressure bump, as overestimating particle growth leads to overestimating pebble trapping.

\subsection{Collisions with two types of material}\label{sec:dust}

Pebbles can consist of refractory inclusions and chondrules embedded in a matrix, forming chondritic aggregates. For each representative particle, we track the relative abundance of rigid (refractory inclusions/chondrules) and fragile (matrix) material. The corresponding rigid mass fraction of particle $i$ is defined as:
\begin{equation}
f^{(d)}_{i} = \frac{m_{ rigid, i}}{m_{i}}\,,
\end{equation}
where $m_{{rigid, i}}$ and $m_{i}$ are the mass of rigid components and the total mass of the particle. The superscript $d$ in $f^{(d)}_{i}$ indicates that the property is dynamic and is updated in every collision. Assuming that the fragile and rigid materials have internal densities of $\rho_{\rm{fragile}} $$\,=$$\, 1.2\,\mathrm{g\,cm^{-3}}$ and $\rho_{\rm{rigid}} $$\,=$$\, 3.3\,\mathrm{g\,cm^{-3}}$ \citep{Hellmann2020}, the internal density of the particle is calculated as:
\begin{equation} \label{eq:rhoi}
    \rho_{i} = \left( \frac{f^{(d)}_{i}}{\rho_{\rm{rigid}}} + \frac{1-f^{(d)}_{i}}{\rho_{\rm{fragile}}} \right)^{-1}\,\,.
\end{equation}

If representative particle $i$ collides with a physical particle $j$ at a relative velocity $\Delta v_{ij}$ lower than the fragmentation threshold $v_{\rm frag}$, and if at least one of the particles contains matrix material, we assume that the two particles stick together. During a sticky collision, the rigid mass fraction of particle $i$ is updated as follows:
\begin{equation}\label{eq:fri_update}
    f^{(d)}_{i} = \frac{f^{(d)}_{i} m_{i} + f^{(d)}_{j}m_{j}}{m_{i}+m_{j}}\,.
\end{equation}
We assume that at typical relative velocities in the disk midplane ($>\mathrm{mm\,s^{-1}}$) two purely rigid particles always bounce when they collide \citep{Gunkelmann2017}.

By tracking only the dynamic fraction $f^{(d)}_{i}$ of two types of material, the Monte Carlo method presented by \cite{Zsom2008} does not guarantee the conservation of the total mass for rigid and fragile material because we only update representative particle $i$ when particle $i$ and $j$ collide \cite[see discussion in ][]{Krijt2016, Houge2023}. This can lead to fluctuations that can grow over time. To address this issue, we introduce static properties for each particle, which specify the type of material to be tracked in a disruptive collision. We define two key properties: $f^{(s)}_{i}$ and $m^{(s)}_{rigid, i}$. Here, the superscript $s$ denotes a static property. If a traced particle is initialized with $f^{(s)}_{i}$$\,=$$\, 1$, it requires a corresponding mass value $m^{(s)}_{rigid, i}$, which is assigned according to a power-law distribution (see Section~\ref{sec:methods_0d}). Throughout the evolution of particle $i$, we consistently track a rigid particle of mass $m^{(s)}_{rigid, i}$. This rigid particle may be embedded within a larger chondritic aggregate containing both types of material. In a disruptive collision, if the tracked fragment cannot retain the rigid monomer of mass $m^{(s)}_{rigid, i}$, we always choose to track the rigid monomer. Conversely, if $f^{(s)}_i = 0$, we track a particle that contains at least one fragile monomer. To ensure the same threshold for separating material regardless of the particle’s $f^{(s)}_i$ value, we track a purely fragile particle when the fragment cannot retain a rigid particle of mass $m^{(s)}_{rigid, i}$. Therefore, if a particle has $f^{(s)}_i = 0$, the static mass $m^{(s)}_{rigid, i}$ is updated whenever the particle collides with a physical particle of similar mass (within 90\%) that has $f^{(s)}_i = 1$. Fragmentation events correct the fluctuations that arise during coagulation. As a result, when the particle size distribution is limited by fragmentation, as in a pressure bump, the dynamic properties fluctuate around their average static values. In this way, fluctuations do not grow over time and decrease with increasing numerical resolution (see Figure~\ref{fig:mass_conservation}).

When representative particle $i$ collides with a physical particle $j$ at a relative velocity $\Delta v_{ij} > v_{\rm frag}$ and particle $i$ contains matrix material, we assume that particle $i$ fragments. The mass of particle $i$ is distributed into fragments following a power-law distribution, and the size of the fragment to be tracked is then selected from the distribution:
\begin{equation}\label{eq:frag1}
    m^{{\prime}}_{i} = \left( \chi \cdot (m^{\kappa}_{i}- m^{\kappa}_{\rm{0}}) +  m^{\kappa}_{0}\right)^{1/\kappa}\,,
\end{equation}
where $\kappa $$\,=$$\, 1/6$ \citep{Dohnanyi1969}, $m_{0}$ is the mass of a fragile monomer of $1\,\mathrm{\mu m}$ in radius, $\chi$ is a random number drawn from a uniform distribution, and $m^{{\prime}}_{i}$ is the mass of particle $i$ after the disruptive collision. We then check if the static rigid monomer in the aggregate ($m^{(s)}_{\mathrm{rigid}, i}$) could be retained in that fragment of mass $ m^{{\prime}}_{i}$. If so, we update the mass while keeping $f^{(d)}_{i}$ constant. This approach is motivated by simulations showing that the chondrules tend to remain in the largest fragments \citep{Umstatter2021}. If the rigid monomer cannot be retained, we check the static property $f^{(s)}_{i}$. If $f^{(s)}_{i}$$\,=$$\,1$, we trace a rigid particle with mass $m^{(s)}_{rigid, i}$. If $f^{(s)}_{i}$$\,=$$\,0$, we trace a fragile particle. We select the mass of the fragile fragment from the power-law distribution in Equation \ref{eq:frag1}, but replacing $m_{i}$ with $ m^{(s)}_{\rm{rigid,i}}/f^{(d)}_{i}$. After choosing the tracked fragment, the dynamic fraction is also updated accordingly.

Laboratory experiments reported a fragmentation velocity threshold of $1\,\mathrm{m\,s^{-1}}$ \citep{Guttler2010}. However, these experiments were conducted with a single material composed of micron-sized monomers. Alternative laboratory experiments with rigid and fragile materials, as well as granular mechanics simulations, have shown that growth is enhanced when both materials are combined \citep{Beitz2012, Gunkelmann2017}. The fragmentation velocity of chondritic aggregates remains uncertain. We assume a constant fragmentation threshold of $v_{\rm{frag}} $$\,=$$\, 2\,\mathrm{m\,s^{-1}}$, both for the reason mentioned earlier and because the Stokes number ($\rm St$, a parameter describing the dust response to aerodynamic drag), limited by fragmentation in a pressure bump, must exceed $0.01$ to trigger planetesimal formation \citep{Bai2010}, which is not possible in our models if we set $v_{\rm{frag}} $$\,=$$\, 1\,\mathrm{m\,s^{-1}}$. We assume that rigid components survive without fragmenting to collisional velocities driven by typical nebular conditions, such as turbulence. Although chondrule fragments are present in chondrites \citep{Scott2014}, most chondrules are preserved as near‑spherical igneous droplets. The presence of fragments may reflect energetic events, such as nebular shocks, which can generate collision velocities high enough to fragment chondrules \citep[e.g.,][]{Ciesla2006, Jacquet2014, Arakawa2019}. Such events are not included in our model.

Matrix in chondrites is commonly found as fine‑grained rims surrounding chondrules and refractory inclusions, as well as interstitial material between them \citep[e.g., ][]{Scott2014, Simon2015}. \cite{Ormel2008} showed that micrometer‑sized matrix material can first stick to chondrules to form porous rims, and that subsequent collisions lead to rim compaction. Afterwards, rimmed chondrules may collide and stick with other rimmed chondrules. In contrast, collisions between rimmed chondrules at relative velocities on the order of $\mathrm{m\,s^{-1}}$ can lead to ejection of rims \citep{Umstatter2019}. Within dust traps, collisions between small matrix and rigid monomers are frequent, and dust collides at relative velocities reaching up to $\sim\,2\,\mathrm{m\,s^{-1}}$ for our disk parameters. Under such conditions, chondrule (and potentially refractory inclusion) rims are likely to form and be disrupted continuously within the same pressure bump. Therefore we simplify the model by assuming perfect sticking between rigid components and matrix material at collision velocities below $2\,\mathrm{m\,s^{-1}}$.

\subsection{Size distribution from zero-dimensional simulations}\label{sec:methods_0d}
We first conduct a zero-dimensional Monte Carlo simulation to analyze the effect of gas removal on the particle size distribution when incorporating the new fragmentation model. The results are presented in Section~\ref{sec:results_0d}. We compute the collisional evolution of a single grid cell with $200$ particles in the dust trap, which for the fiducial disk parameters at $2\,\mathrm{Myr}$ is located at $6.1\,\rm{AU}$. We assume a vertically integrated dust-to-gas ratio of $0.05$ for this simulation. This metallicity is consistent with the value expected to trigger planetesimal formation for the Stokes number $\rm{St}$$\,=$$\,0.01$ (Equation~\ref{eq:Lim}).

We initialize particles using a power-law size-distribution consistent with fragmentation-limited dust growth, $ n(a) $$\propto$$\, a^{-3.5} $ \citep{Birnstiel2024}, which is later stabilized by the simulation. We randomly choose half of the particles to have $f_{\rm{s, rigid}}$$\,=$$\,1$. We determine the mass of the rigid monomer ($m_{\rm{s, rigid}}$) from the power-law distribution $n(a)$$\,\propto$$\,a^{\zeta}$, assuming spherical monomers with radii $a$. We took $10\,\mathrm{\mu m}$ and $1000\,\mathrm{\mu m}$  as the minimum and maximum sizes of the distribution, and a power index of $\zeta$$\,=$$\,-3.9$, resulting in a mean radius of $230\,\mathrm{\mu m}$. These values cover the typical sizes of CAIs and chondrules \citep{Jones2012, Dunham2023}.

After initializing the static properties, we select particle masses from the fragmentation-limited power-law distribution. We then check whether a particle of the chosen size could retain $ 50\,\%$ of its mass in rigid monomers. If this condition is met, the particle is initialized with $f_{\rm{d, rigid}}$$\,=$$\,0.5$. For particles that are too small to contain rigid monomers, when $f_{\rm{s, rigid}} $$\,=$$\, 0$, the particles are chosen to be pure fragile with the previously chosen mass. Otherwise, if $f_{\rm{s, rigid}}$$\,=$$\,1$, the particle is initialized as a rigid monomer with mass $m_{\rm{s, rigid}}$.

To illustrate the particle size distribution, in Figure\,\ref{fig:0D} we compute the surface density on a logarithmic scale as
\begin{equation}\label{eq:sigmad}
    \sigma_{\rm{d}}(a) = N (a) m(a) a\,,
\end{equation}
where $a$ is the radii of dust particles, $m$ the particle mass and $N$ the number of particles with the corresponding radius \citep{Birnstiel2024}.

\subsection{Global simulations of dust evolution}\label{sec:feeding}

We then perform simulations that include dust collisional evolution, transport, and planetesimal formation. Due to the high computational demand of the Monte Carlo simulations, we divide our simulations into two parts. We first ran a global dust evolution simulation, starting at $2\,\mathrm{Myr}$ after CAI formation, to model the delivery of pebbles to the dust trap. We compute the collisional evolution and transport in the outer regions of the disk ($r$$\,>$$\,6.5\,\mathrm{AU}$). Later, we use the output of this simulation to calculate the pebble flux at $6.5\,\mathrm{AU}$ in a local simulation of the outer gap edge (see next section).

We assume an integrated dust-to-gas ratio of $Z_{0}$$\,=$$\,5$$\,\cdot $$\,10^{-5}$ across a $2\,\mathrm{Myr}$-old disk, corresponding to a total dust mass of approximately $1.2$ Earth masses. This estimate aligns with the low dust masses observed in Class II disks \citep{Ansdell2016}. We initialize the particle size distribution similarly to the previous zero-dimensional simulations. However, instead of assuming a uniform abundance of rigid material, we decrease the rigid abundance with radial distance. Assuming that the largest rigid particles carry most of the rigid mass, their decreasing abundance with radial distance is consistent with dust evolution models, which predict that millimeter-sized grains undergo significant radial drift before $2\,\mathrm{Myr}$ in the outer disk. On the other hand, because we adopt a size distribution for the rigid monomers (ranging from $10\,\mathrm{\mu m}$ to $1\,\mathrm{mm}$) and allow the matrix to grow into large pebbles, the reduction of rigid abundance in the outer disk is expected to be less pronounced than in models that assume fixed particle sizes and millimeter-sized refractory inclusions \citep[e.g.,][]{Desch18, Jongejan2023}.

To model the initial decrease of rigid particles with distance, we first set an initial value of $\bar{f}_{\rm{rigid}}$, and then the value is modified by multiplying it by:
\begin{equation}
    p_{\rm{rigid}} = \frac{(a_{\rm{ri, lim}})^{\zeta+4} -  (a^{\rm{min}}_{\rm{ rigid}})^{\zeta+4}}{(a^{\rm{max}}_{\rm{ rigid}})^{\zeta+4} -  (a^{\rm{min}}_{\rm{ rigid}})^{\zeta+4}}\,,
\end{equation}
where $a_{\rm{ri, lim}}$ is the size limited by fragmentation or radial drift, assuming the Epstein regime \citep{Birnstiel2024}, given by:
\begin{equation} \label{eq:mrilim}
     a_{\rm{ri, lim}} = \frac{2}{\pi} \frac{\Sigma_{\rm{g}}\rm{St_{lim}}}{\rho_{\rm{rigid}}} \,,
\end{equation}
and $a^{\rm{min}}_{\rm{rigid}}$ is $10\,\mathrm{\mu m}$ and $a^{\rm{max}}_{\rm{rigid}}$ is the minimum value between $1000\,\mathrm{\mu m}$ and $a_{\rm{ri, lim}}$. $\rm{St}_{lim}$ is the Stokes number limited by fragmentation or radial drift \citep{Birnstiel2012}, and therefore the minimum value between
\begin{equation} \label{eq:Stfrag}
    \mathrm{St_{frag}} =\frac{0.37}{3\alpha_{\rm{t}}}\left( \frac{v_{\rm{frag}}}{c_{\rm{s}}}\right)^{2}\,,
\end{equation}
\begin{equation}\label{eq:Stdrift}
    \mathrm{St_{drift}} =\frac{1}{2} \frac{v_{\rm{K}}}{\Delta v}Z_{0}\,,
\end{equation}
where $v_{\rm{K}}$ is the Keplerian velocity and $\Delta v$ is the deviation from the Keplerian rotation of the gas (see Equation \ref{eq:deltav}). In regions where dust growth is limited by the drift barrier (Equation~\ref{eq:Stdrift}), the initial particle size distribution follows $ n(a)$$\,\propto$$\,a^{-2.5}$ \citep{Birnstiel2024}. The probability of initiating a rigid particle is then given by $\bar{f}_{\rm{rigid}}\times p_{\rm{rigid}}$, and the dynamic rigid mass fraction of particles ($f_{\rm{d, rigid}}$) is set to the same value as this corrected probability. The abundance of rigid material in the solar system’s protoplanetary disk is uncertain. We set $\bar{f}_{\rm{rigid}}$$\,=$$\,0.5$, resulting in an overall rigid disk mass of $0.3$ Earth masses, approximately $25\,\%$ of the total dust mass, and an average rigid monomer radius of $260\,\mathrm{\mu m}$. These values decrease with distance from the Sun; for instance, beyond $50\,\mathrm{AU}$, the overall rigid disk mass drops to $\sim 15\%$, with an average rigid monomer radius of $45\,\mathrm{\mu m}$. With this assumption, our results are consistent with meteorite constraints (see Section~\ref{sec:discussion}). More importantly, the dynamical processes we identify in this work are independent of this choice.

We simulate 80 radial and 40 vertical grid bins, each cell containing 200 particles. No feedback from local to global simulations is included; once particles cross the inner boundary of the global simulation, they are no longer tracked. To stabilize the collisional model, we first perform 100 iterations without transport. Then, we simulate dust evolution until the gas disk dissipates, approximately $4.2\,\mathrm{Myr}$ after CAI formation (see Figure \ref{fig:gas}). The output of this simulation is later used as the source of a more localized simulation. All parameters employed to calculate dust evolution are listed in Table~\ref{tab:dust}.

\subsection{Local simulation with planetesimal formation}\label{sec:local}
We simulate dust evolution at the outer edge of the gap with a resolution high enough to resolve planetesimal formation. For this setup, we chose an inner boundary of $5.725\,\mathrm{AU}$ to resolve collisions, as particles that cross this location commonly leak through the gap without colliding due to their high radial velocities driven by advection. We set the outer boundary at $6.5\,\mathrm{AU}$ for a dust trap located at $6.1\,\mathrm{AU}$.

We employ a grid with 28 radial and 28 vertical cells, with each cell containing 200 particles, as the simulation converges for this resolution. Particles are introduced at $6.5\,\mathrm{AU}$ from the output of the global simulation, while some particles are lost as they pass through the gap. As the simulation progresses, the number of particles naturally would increase, making it computationally challenging to extend the run until $4.2\,\mathrm{Myr}$. To address this issue, we maintain a constant number of particles throughout the simulation. Then, the total mass of each representative particle ($M_{\rm{swarm}}$), which is the same for all particles, changes over time. We randomly remove or duplicate particles as needed to keep the number of particles constant. This approach allows us to extend the simulation over long timescales without increasing computational costs at later times. We validate our results by running the simulation multiple times with the same initial conditions and confirming their consistency.

We implement a simple criterion for simulating planetesimal formation by the Streaming Instability (SI): if the integrated pebble-to-gas ratio exceeds $Z_{\rm{crit}}$ from \cite{Lim2024} (their Equation 19), 
\begin{equation}\label{eq:Lim}
\begin{aligned}
\log Z_{\rm crit} =\;& 0.15\,\log^{2}(\alpha_{\rm t})
 - 0.24\,\log(\mathrm{St})\,\log(\alpha_{\rm t}) \\
& - 1.48\,\log(\mathrm{St})
 + 1.18\,\log(\alpha_{\rm t}) \,,
\end{aligned}
\end{equation}
we trigger planetesimal formation. This expression was derived for pebbles with a single Stokes number in the range $0.01 \leq \rm{St} \leq 0.1$, whereas our simulations include particle size distribution. For our fiducial parameters, the Stokes number limited by fragmentation is near 0.02 (Equation \ref{eq:Stfrag}). To calculate the critical metallicity for SI, we select the lower value of $\rm{St}=0.01$, which, together with $\alpha_{\rm{t}}$$\,=$$\,10^{-4}$, yield a value of $Z_{\rm{crit}}=0.05$. As the gas disperses, rigid particles acquire Stokes numbers that exceed the fragmentation limit (see Section~\ref{sec:results_0d}). While we expect that pebbles with $\rm{St} > 0.1$ can still form planetesimals, the upper Stokes number limit remains uncertain. We set this limit at $\rm{St}$$\,=$$\,1$, beyond which solids no longer behave as pebbles. Nevertheless, particles that trigger the SI remain below a Stokes number of 0.1 throughout our simulations.

If the criterion for forming planetesimals is fulfilled, we compute the number of representative particles converted into planetesimals during the timestep $dt$ following:
\begin{equation}\label{eq:plts_formation}
     N_{\rm{plt}} = \zeta_{\rm{eff}} \frac{N_{\rm{swarm}}(\rm{St}>\rm{St_{crit}})}{T_{\rm{K}}}dt\,,
\end{equation}
where $T_{\rm{K}}$ is the orbital period and $\zeta_{\rm{eff}}$ denotes the planetesimal formation efficiency \citep{Drazkowska2016}. We adopt a fiducial value of $\zeta_{\rm{eff}} $$\,=$$\, 10^{-3}$ \citep{Simon2016}. We assume that the planetesimals have the same composition as their pebble precursors.
 
\begin{figure*}
\centering
\includegraphics[width=0.7\textwidth]{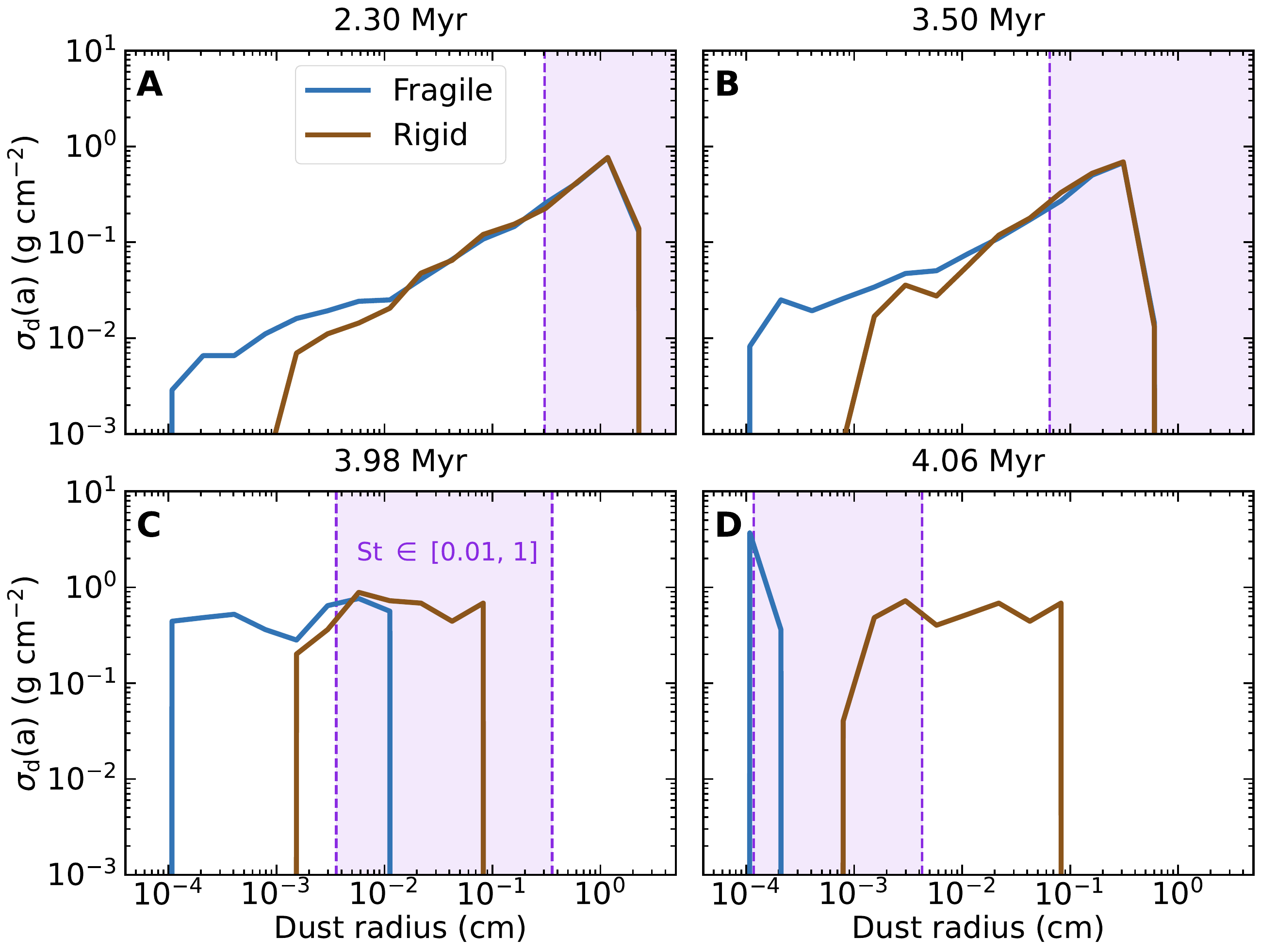}

\caption{Particle size distribution from solely collisional evolution of fragile and rigid material at different stages of disk evolution in a zero-dimensional simulation. The surface density is computed on a logarithmic scale $\sigma_{\rm{d}}$ (see Equation~\ref{eq:sigmad}). Panels (\textbf{A}) and (\textbf{B}) show time-averaged results over 50 snapshots, while the panels (\textbf{C}) and (\textbf{D}) represent single-time snapshots. The total mass is equally distributed between 200 fragile and rigid particles. Rigid monomer radii range from $10$ to $1000\,\mathrm{\mu m}$, while fragile monomers are $1\,\mathrm{\mu m}$.  The purple dashed lines delimit the domain of interest for planetesimal formation, for pebbles with
Stokes numbers ($\rm St$) between 0.01 and 1 for an internal density of $1.76\,\mathrm{g\,cm^{-3}}$ (corresponding to a rigid mass fraction of 0.5), except in the panel (\textbf{D}), where the rigid mass fractions are 0 and 1 for Stokes numbers of 0.01 and 1, respectively. $\rm{St}$$\,=\,$$1$ line is not displayed in panels (\textbf{A}) and (\textbf{B}) because it corresponds to particle sizes larger than those shown. (\textbf{A})-(\textbf{C}) The maximum aggregate size of combined fragile and rigid material decreases over time due to ongoing fragmentation. (\textbf{D}) At the latest stages, the chondritic aggregates are fragmented into their constituent components.}\label{fig:0D}

\end{figure*}

\section{Results}\label{sec:results}

\begin{figure*}
\centering
\includegraphics[width=0.9\textwidth]{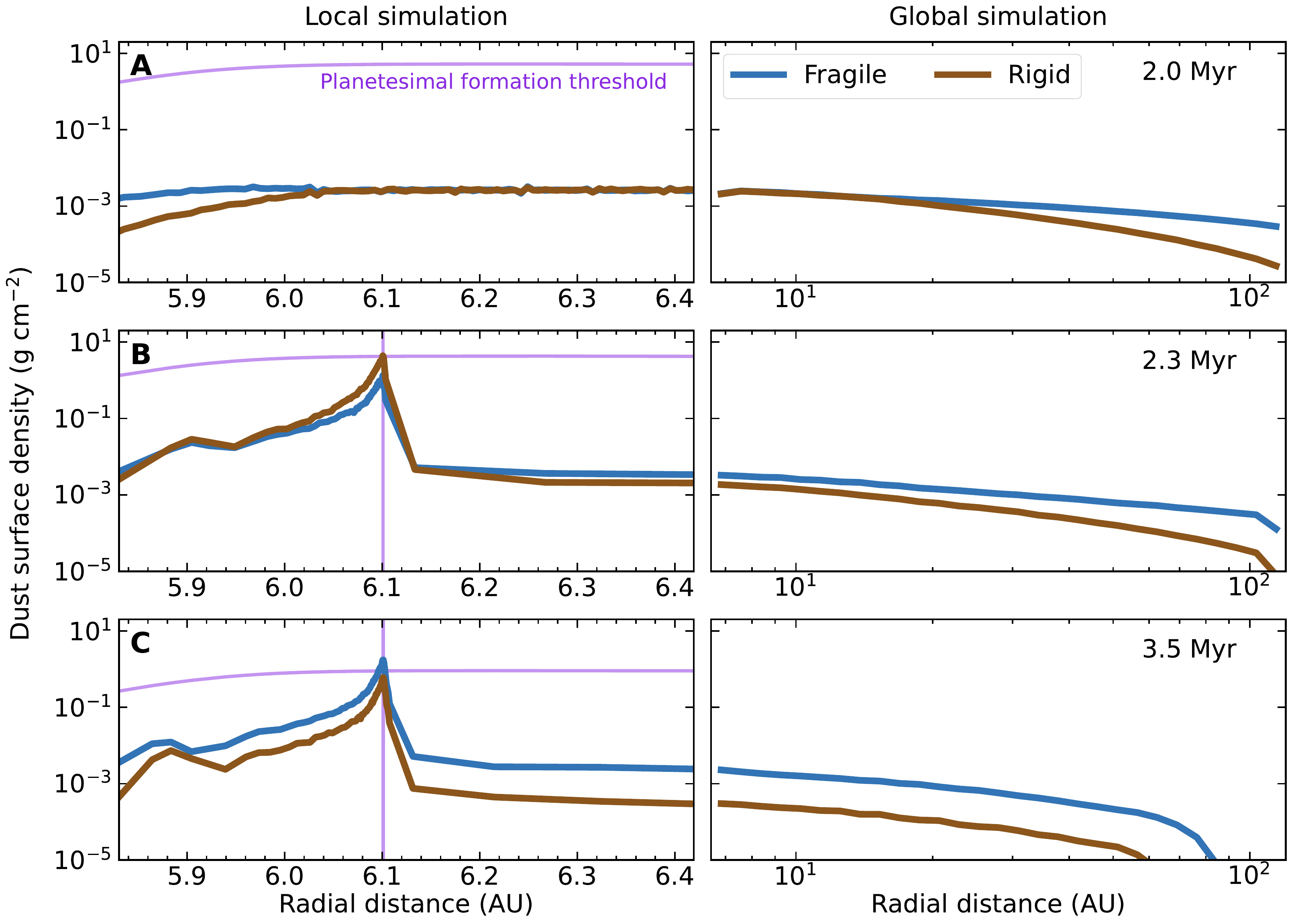}
\caption{Dust surface density evolution across the disk. Local and global simulations are shown together at different snapshots. (\textbf{A}) Start of the simulation at $2\,\mathrm{Myr}$. Dust is distributed equally between rigid and fragile material, except when restricted by the radial drift barrier. The purple (nearly) horizontal line indicates the minimum dust surface density required for planetesimal formation. (\textbf{B}) Simulation at $2.3\,\mathrm{Myr}$. The dust surface density in the dust trap increases, particularly for rigid material due to the leakage of fragile material. The purple vertical line indicates the location where planetesimals form at approximately $6.1\,\mathrm{AU}$. (\textbf{C}) Simulation at $3.5\,\mathrm{Myr}$. The surface density of rigid material in the trap decreases, as the material replenishing the dust trap is matrix-rich.}\label{fig:local}
\end{figure*}

\subsection{Collisional evolution of dust with different fragility}\label{sec:results_0d}

To understand how including fragile and rigid particles affects dust evolution, we first perform a zero-dimensional simulation, in which we distribute half of the material as fragile and the other half as rigid monomers. Fragile material is composed of monomers with radii of $1\,\mathrm{\mu m}$ and a density of $1.2\,\mathrm{g\,\mathrm{cm^{-3}}}$, whereas rigid monomers have radii ranging from $10$ to $1000\,\mathrm{\mu m}$ and are characterized by a higher density of $3.3\,\mathrm{g\,cm^{-3}}$. All dust particles stick to each other upon collision if the impact velocity remains below the fragmentation threshold of $2\,\mathrm{m\,s^{-1}}$, except for collisions between rigid monomers, which result in bouncing (see Section~\ref{sec:dust} for details of the fragmentation model and Section~\ref{sec:methods_0d} for a description of the numerical implementation).

When the dust grows to centimeter and millimeter-sized chondritic aggregates (Figures~\ref{fig:0D}A and \ref{fig:0D}B), the largest pebbles are composed of rigid and fragile material equally. Most pebbles have Stokes numbers between 0.01 and 1, and thereby can trigger planetesimal formation by the SI if the pebble-to-gas ratio is high enough. When the gas surface density decreases significantly by photoevaporation, the sizes at which aggregates containing both fragile and rigid material shift to sub-millimeter sizes (Figure~\ref{fig:0D}C). At the same time, larger rigid particles remain as monomers with Stokes numbers larger than $0.01$. When the gas disk depletes even further (Figure~\ref{fig:0D}D), many of the rigid monomers are too large to trigger planetesimal formation ($\rm{St}\,$$>\,$$1$), while the Stokes numbers of matrix particles lie in the optimal range for planetesimal formation. These results will be important for understanding the late formation of the CR and CI chondrite parent bodies in the next section.

\subsection{Dust evolution across the disk}

\begin{figure*}
\centering
\includegraphics[width=0.9\textwidth]{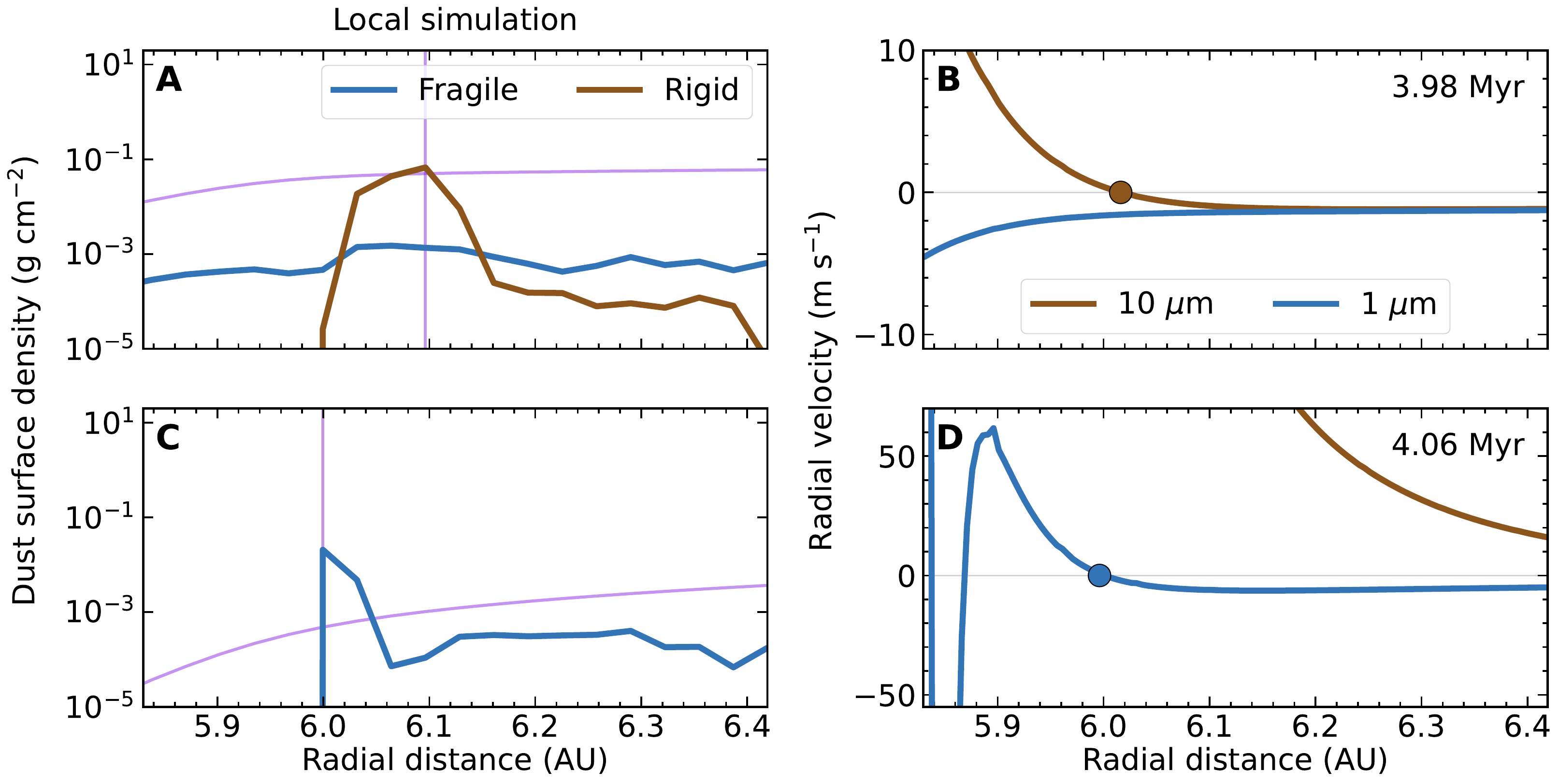}
\caption{Dust evolution at the outer edge of the gap during the gap-widening phase driven by photoevaporation. (\textbf{A}) Dust surface density at $3.98\,\mathrm{Myr}$. The purple (nearly) horizontal line indicates the minimum dust surface density required for planetesimal formation, and the vertical one the location where this threshold is exceeded. (\textbf{B}) Radial velocities of the smallest rigid and fragile monomers at $3.98\,\mathrm{Myr}$. The dot at around $6\,\mathrm{AU}$ denotes the location where the smallest monomers of each material get trapped, corresponding to where their radial velocity is zero. At this time, small matrix fragments leak through the gap along with the gas flow, while all rigid particles remain trapped. (\textbf{C}) Dust surface density and (\textbf{D}) radial velocities at $4.06\,\mathrm{Myr}$. By this time, rigid monomers have been pushed outward by the expanding gap, while the smallest matrix monomers are now trapped near $6\,\mathrm{AU}$. This sequence explains the dynamics during the formation of matrix-poor planetesimals and the subsequent formation of matrix-rich ones.}\label{fig:CRCI}
\end{figure*}

While the simulation described above highlights the critical role of the gas evolution in shaping the dust size distribution, a realistic model must also account for dust filtering, material delivery from outer regions, and planetesimal formation. To model how the dust population in the disk evolved in time and space, we ran a global simulation of the outer disk. The radial drift velocity of particles scales with their Stokes number, which increases with particle size and in regions of lower gas density, such as the outer disk \citep{Birnstiel2024}. As a result, millimeter-sized rigid particles in the outer disk drift inward on timescales shorter than a $\rm{Myr}$. To initialize a Class II disk, we reduce the dust-to-gas ratio and the abundance of large rigid monomers in the outer disk to reflect their earlier rapid inward drift (see Section~\ref{sec:feeding}). This results in an outer disk that is relatively enriched in fragile material (i.e., fine-grained matrix). We then run a high-resolution local simulation at the outer edge of the gap (see Section~\ref{sec:local}). When the conditions for planetesimal formation are satisfied, we remove the pebbles that became part of planetesimals and assume that these planetesimals inherit the same rigid mass fraction as their pebble precursors.

Figure \ref{fig:local} shows the dust evolution in the disk for rigid and fragile materials. Initially, these materials are equally distributed across the disk (Figure~\ref{fig:local}A), except in the outer disk, where, as noted above, the abundance of rigid material is reduced due to the fast radial drift of large rigid monomers. By $2.3\,\mathrm{Myr}$, pebble trapping at the outer edge of the gap increases the dust density around $6.1\,\mathrm{AU}$ for both rigid and fragile material (Figure~\ref{fig:local}B), but the increase is higher for rigid than for fragile material. This is because collisions between aggregates are frequent in the dust trap, and high-velocity impacts tend to fragment fragile aggregates more efficiently into smaller dust particles. These smaller fragments are more strongly coupled to the gas and can leak through the gap, while the larger and denser rigid monomers remain trapped. At around $2.3\,\mathrm{Myr}$, the total dust surface density exceeds the threshold for triggering planetesimal formation (Figure~\ref{fig:local}B). By $3.5\,\mathrm{Myr}$, the dust surface density from rigid material has decreased significantly with respect to that of the fragile material (Figure~\ref{fig:local}C). This decrease is primarily due to a temporal reduction in the number of rigid particles delivered to the dust trap while planetesimal formation regulates the dust amount in the trap (see Figure~\ref{fig:differente}). While pebbles are incorporated into planetesimals over time, the dust trap is replenished by material that drifts from the outer regions. By this time, sub-millimeter-sized rigid particles, which are the main carriers of the total rigid mass, have already drifted into the dust trap. As a result, the dust population is now replenished mainly by fragile matrix-like material. Consequently, the differences in dust filtering and delivery rates of rigid and fragile material can account for the low abundance of matrix in early-formed carbonaceous chondrites and the increase in matrix in later-formed ones.

At the later stage of disk evolution, at around $4\,\mathrm{Myr}$, photoevaporation begins to widen the existing gap (see Figure~\ref{fig:gas}). As gas is expelled from this region, the surrounding gas attempts to refill the gap, increasing the gas velocity near the outer edge of the gap. Micrometer-sized fragile particles, which remain well coupled to the gas, move along with the gas in an effort to refill the gap. As a result, they are carried away from the planetesimal formation region (Figure~\ref{fig:CRCI}A). In contrast, rigid monomers, which are denser and larger, are efficiently trapped in the dust trap (Figure~\ref{fig:CRCI}A) and have high Stokes numbers (Figure~\ref{fig:0D}C). This increases the ratio of rigid-to-fragile material at the outer edge of the gap (Figure~\ref{fig:CRCI}A). If sufficient similar-sized rigid particles are present, they can form planetesimals composed of matrix-poor dust, reminiscent of the CR chondrites. At even later times (Figure~\ref{fig:CRCI}B), the gap widens and the gas density decreases further. By this point, most rigid particles have either been incorporated into planetesimals or transported outward. Simultaneously, the smallest fragile monomers begin to accumulate in the trap, leading to the formation of planetesimals made primarily of matrix, such as the CI chondrites. Together, these results show that photoevaporation can lead to the late formation of matrix-poor planetesimals from mostly rigid monomers, followed by the formation of matrix-rich planetesimals.

\section{Discussion}\label{sec:discussion}
\subsection{Prolonged carbonaceous chondrite formation in a single pressure bump}

\begin{figure}
\centering
\includegraphics[width=0.45\textwidth]{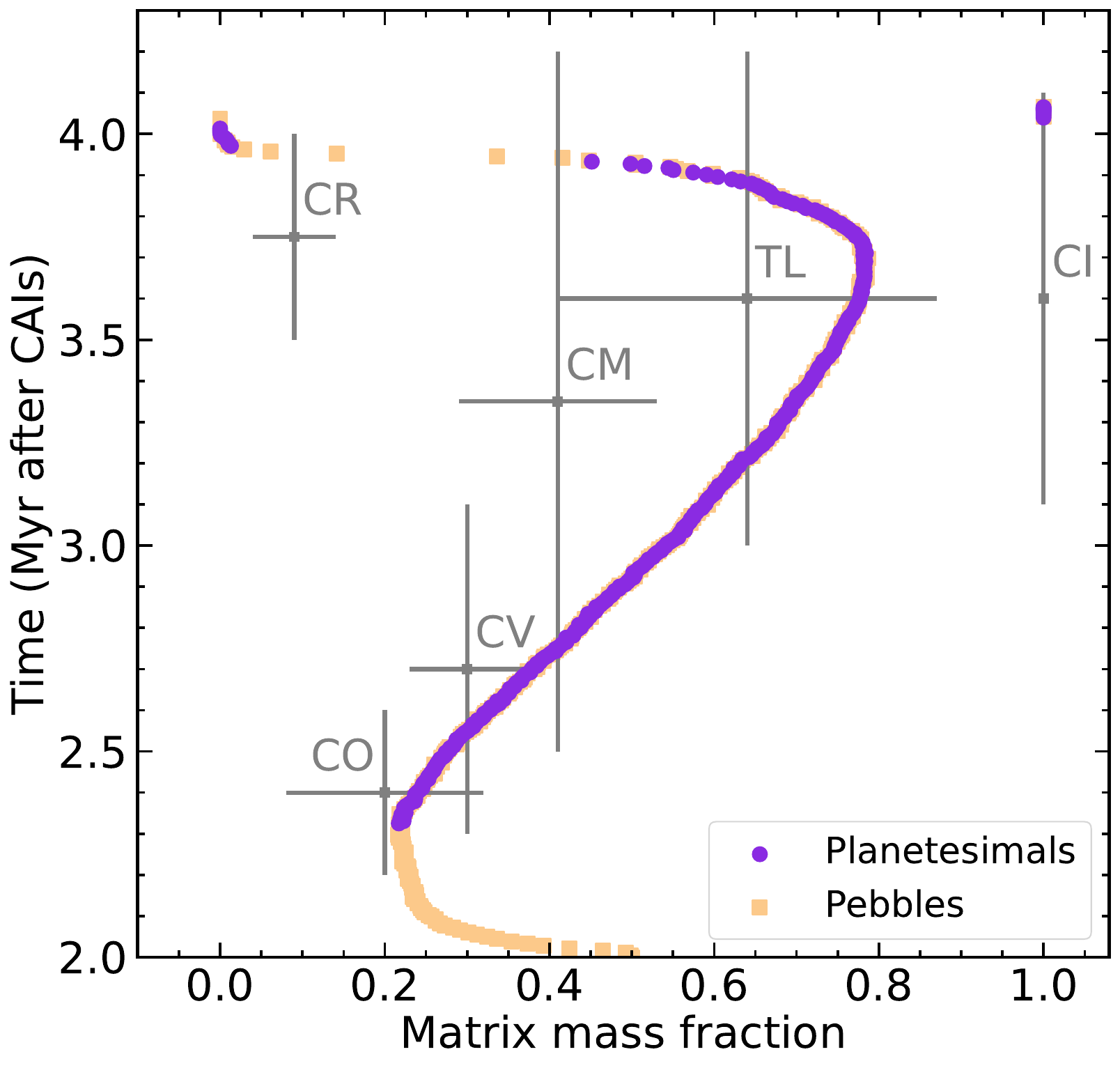}
\caption{Matrix mass fraction in planetesimals depending on their formation time. Planetesimals are indicated with purple dots and the largest pebbles in the dust trap (with Stokes numbers between 0.01 and 1) with orange squares. Meteoritic data represented with gray points with error bars are from \cite{Hellmann2023} and references therein.}\label{fig:plt}
\end{figure}

The key observation from our results is that they fully reproduce the observed compositional range of carbonaceous chondrite parent bodies over time. This is illustrated in Figure~\ref{fig:plt}, where we show the temporal evolution of the matrix mass fraction of planetesimals and pebbles in the dust trap from our model compared to the observed ages and compositions of carbonaceous chondrites. Pebbles start with an assumed average matrix mass fraction of $50\,\%$ at $2\,\mathrm{Myr}$. This fraction initially decreases due to the loss of the matrix passing through the gap, which halts at a matrix fraction of around $20\%$ at $\sim\,$$2.3\,\mathrm{Myr}$. At this point, the conditions for planetesimal formation are met for the first time, reproducing the accretion age and composition of the Ornans- (CO) and Vigarano-type (CV) chondrites. Since the dust delivered from the outer disk is more matrix-rich, the matrix fraction of pebbles and planetesimals increases until $\sim$$\,$$3.8\,\mathrm{Myr}$, reproducing the properties of the more matrix-rich Mighei-type (CM) chondrites and the ungrouped Tagish Lake (TL) chondrite. It is important to note that incorporating multi-component dust evolution and planetesimal formation, as in this study, is required to quantify this temporal increase of matrix in the pressure bump (see Figure \ref{fig:differente}). By contrast, previous studies that did not consider dust growth and planetesimal formation have shown that the abundance of refractory inclusions in the pressure bump increases over time (\cite{Desch18} their Figure~8, and \cite{Jongejan2023} their Figure~17), which is contrary to the observations among carbonaceous chondrites.

As the protoplanetary disk approached its dispersal, the last phase of planetesimal formation occurred in two stages: first, remnant rigid particles formed CR-like planetesimals and, later, the smaller, less dense fragile grains formed CI-like planetesimals (Figure~\ref{fig:CRCI}). In the first stage, the matrix abundance in the largest pebbles and planetesimals rapidly decreases due to the gap widening during photoevaporation (Figure ~\ref{fig:CRCI}A and \ref{fig:CRCI}B) and, as such, can reproduce the very low abundance of matrix in the CR chondrites. Forming CR-like planetesimals in this manner requires that enough rigid particles remained at around $4\,\mathrm{Myr}$ for a local pebble-to-gas ratio high enough for planetesimal formation. Our simulations reproduce this scenario, forming matrix-poor bodies, but we also found scenarios in which planetesimal formation is not triggered due to the scarcity of rigid particles at a later stage (see Figure~\ref{fig:CR_formation}). However, these scenarios can still explain the formation of CR chondrites if a later generation of chondrules formed, supplying the rigid material required for planetesimal formation. This interpretation is consistent with the distinctive composition of CR chondrites, which are rich in late-formed chondrules but poor in refractory inclusions \citep{Budde2018, Bryson2021, Marrocchi2022}.

Like the CR chondrites, CI chondrite-like planetesimals also form late, and our results suggest their formation times differ by less than $0.1\,\mathrm{Myr}$ (Figures~\ref{fig:CRCI} and \ref{fig:plt}). The CI chondrites formed when only micrometer-sized fragile particles are trapped and meet the criteria for planetesimal formation (Figure~\ref{fig:CRCI}C and \ref{fig:CRCI}D), reproducing the high matrix abundance of these carbonaceous chondrites. In our model, planetesimal formation is only monitored at the outer gap edge of the local simulation, but we expect that matrix-rich planetesimals continue to form across the outer disk as the gap expands \citep{Carrera2017, Ercolano2017}. The CI chondrites are isotopically distinct in particular for Fe and Ni, which has been suggested to reflect either formation in a spatially separated region of the disk \citep{Hopp2022} or fractionation of isotopically anomalous FeNi metal grains during carbonaceous chondrite formation \citep{Spitzer2024}. Although FeNi metal grains have not been included in our model, our findings point toward the latter scenario because they show that CI chondrite-like objects can form in the same disk region and from the same starting materials as other carbonaceous chondrites.

In our fiducial simulation, the total mass of planetesimals formed in the pressure bump is approximately $0.5$ Earth masses. Most of the planetesimals are compositionally similar to CO, CV, CM, or TL ($\sim99.2\%$; see Fig.~\ref{fig:planetesimal_mass}), and CR-like and CI-like planetesimals are significantly less abundant ($\sim0.6\%$ and $0.2\%$, respectively). However, these fractions vary under additional considerations. Including late-generation chondrule formation (i.e., the aforementioned supply of rigid material at late stages) may increase the CR fraction. In addition, tracking planetesimal formation across the outer disk as the gap opens raises the CI fraction to $\sim5\%$. Additional refinements in the model, such as adopting a more extended and massive outer disk, would further increase the CI fraction. More work is required to assess whether the mass distribution of carbonaceous chondrite parent bodies predicted by our model is consistent with current observational constraints. On the one hand, the fraction of the formed planetesimals that is later implanted into the asteroid belt, the main source of meteorites \citep{Colas2020}, should be quantified. On the other hand, meteorite collections alone almost certainly do not provide a representative sample of parent body abundances. Notably, although CI chondrites are rare among meteorite falls on Earth, JAXA’s Hayabusa2 mission to asteroid Ryugu and NASA’s OSIRIS-REx mission to asteroid Bennu revealed that these two asteroids are CI-like \citep{Yokoyama2023, Lauretta2024, Barnes2025}, suggesting that this planetesimal group may be more common than inferred from meteorite collections.

In summary, our results show that although carbonaceous chondrites are compositionally distinct and formed over an extended period of time, they can all have formed in a single long-lived dust trap of the disk. If instead photoevaporation were the sole mechanism for enhancing the local pebble-to-gas ratio, planetesimal formation would first occur in the outer disk beyond $100\,\mathrm{AU}$ \citep[][their Figures~6 and 7]{Carrera2017}. These outer-disk planetesimals would be expected to be matrix-rich, as radial drift depletes this region of refractory inclusions and chondrules. Subsequent inside-out photoevaporation could then generate planetesimals at smaller heliocentric distances, but only within a narrow time interval of $\sim\,$$0.1\,\mathrm{Myr}$ \citep{Carrera2017}, with compositions likely similar to those of CI chondrites (Figure \ref{fig:CRCI}). All these predictions are inconsistent with the prolonged period of formation of compositionally distinct bodies as recorded in the carbonaceous chondrites \citep{Hellmann2023}. Our results, therefore, show that a long-lived pressure bump provides a more plausible explanation for the temporal and compositional diversity of carbonaceous chondrites than photoevaporation alone.

\subsection{Implications for early planetesimal formation in the outer disk}

Although our model is devised to examine the formation of carbonaceous chondrites, our results also have implications for understanding earlier planetesimal formation in the outer disk. Iron meteorites sample the metal cores of differentiated planetesimals and, based on their Hf-W core formation ages, are thought to have formed within the first Myr of the solar system \citep{Kleine2005, Kruijer2014}. Thus, iron meteorites represent an earlier generation of planetesimals than the carbonaceous chondrites. Among these irons, the carbonaceous chondrite (CC)-type irons have isotopic compositions similar to those of the carbonaceous chondrites and, therefore, appear to have formed from the same precursor materials as the carbonaceous chondrites \citep{Kruijer2017}. Based on their isotopic compositions, it has been argued that CC irons, like the matrix-poor carbonaceous chondrites, are enriched in refractory inclusions and chondrules, while irons with CI chondrite-like isotopic compositions do not seem to exist \citep{Spitzer2025}. Combined with the results of our model, this suggests that the CC irons, like the carbonaceous chondrites, formed in a dust trap, indicating that such substructures in the disk already existed earlier than modeled here. Importantly, the lack of irons with CI chondrite-like isotopic compositions is consistent with this interpretation, because, as noted above, CI chondrites only formed at the end of the disk’s lifetime, when photoevaporation enhanced the pebble-to-gas ratio and allowed the accretion of micrometer-sized dust components. Since the CC irons formed much earlier, several Myr before photoevaporation occurred, these objects cannot have acquired a CI chondrite-like isotopic composition.

Several other CC iron meteorites exhibit CR chondrite–like isotopic compositions \citep{Spitzer2025}. As argued above, like for the CI chondrites, photoevaporation may have been important in the formation of the CR chondrites, raising the question of how objects with an CR-like isotopic composition can also have formed early, several Myr before photoevaporation commenced. It has been proposed that CR chondrites formed from matrix-rich precursor material, which has been converted into chondrules prior to CR chondrite parent body accretion \citep{Marrocchi2022}. Thus, one possibility is that some CC irons acquired a CR-like isotopic composition by accreting this material predominantly as matrix-rich pebbles. As such, early-formed planetesimals may have undergone an independent sequence of matrix accumulation over time in a substructure (analogous to the progressive matrix increase illustrated in Figure~\ref{fig:local}C). This hypothesized substructure, however, does not necessarily originate from a fully formed Jupiter \citep{Brasser2020, Morbidelli2024}. At earlier times, when the global dust-to-gas ratio was higher, even smaller pressure bumps could have accumulated enough material to trigger planetesimal formation. This pressure bump could have triggered Jupiter's formation \citep{Lau2024}, thus explaining the observed prolonged separation of carbonaceous and non-carbonaceous reservoirs.

\subsection{Implications for the formation of non-carbonaceous chondrites}

Substructures are common in protoplanetary disks \citep{Andrews2018}, and a single planet can create multiple pressure bumps in its inner regions \citep{Lega2025}, raising the question of whether the non-carbonaceous (NC) chondrites, like their carbonaceous counterparts, also formed in a substructure of the disk. The NC chondrites, such as ordinary and enstatite chondrites, formed at around 2 Myr after solar system formation \cite[e.g.,][]{Sugiura2014}, and their isotopic compositions indicate they formed in a different region of the disk than the carbonaceous chondrites, most likely in the inner disk \citep{Kleine2020}. Importantly, the NC chondrites are depleted in refractory inclusions compared to most carbonaceous chondrites \citep{Scott2014}, but are rich in chondrules. Moreover, their chondrules exhibit distinct isotopic signatures than those from carbonaceous chondrites \citep{Schneider2020}. Our results are consistent with these observations, because at $\sim$2 Myr most of the mass in refractory inclusions and (CC-like) chondrules was concentrated in sub-millimeter-sized particles, which leads to their efficient trapping and hence limited transport to the inner disk \cite[see also e.g.,][]{Desch18, Weber2018, Haugbolle2019, Jongejan2023, Clepper2025}. Also, given that substructures in the disk act as traps mostly for rigid material, while more fragile material is transported away, the formation of NC chondrites in such substructures would naturally explain their high (NC-like) chondrule abundance. We thus speculate that the non-carbonaceous and carbonaceous chondrites formed in substructures inside and outside of the same planetary gap.

\section{Conclusions}\label{sec:conclusion}
We have presented a model that integrates gas disk evolution, dust growth, and planetesimal formation while accounting for the different properties of dust components in carbonaceous chondrites. Our simulations show that differences in dust filtering and delivery rates to a planet-induced pressure bump can reproduce the matrix abundances relative to refractory inclusions and chondrules observed in carbonaceous chondrites over time \citep{Hellmann2023}. These results support a scenario in which carbonaceous chondrite parent bodies formed in a single, long-lived pressure bump, most likely located outside Jupiter’s orbit.

In our model, chondrules and refractory inclusions stick to the matrix to form pebble-sized aggregates that can trigger planetesimal formation. Because refractory inclusions and chondrules are larger and more resistant to fragmentation than matrix grains, pressure bumps become enriched in these rigid components relative to matrix. In addition, larger particles drift inward faster, so material arriving from the outer disk becomes progressively more matrix-rich with time. This naturally explains the sequence from matrix-poor planetesimals (CO, CV) to progressively more matrix-rich bodies (CM, TL) in a pressure bump.

At later stages of disk evolution, when the gas density is low, collision velocities become too high for dust growth. In our simulations, matrix-poor planetesimals (analogous to CR chondrites) form from rigid monomers of tens of micrometers, whereas matrix-rich planetesimals form later from micrometer-sized matrix dust. Since during the disk dispersal the planetary gap opens, the formation of these matrix-rich planetesimals could extend to the outer disk.

\section*{Data and code availability} \label{sec:cite}

A curated subset containing all data necessary to interpret, verify, and extend the findings of this study has been deposited in Zenodo, which can be accessed at \href{https://zenodo.org/records/17176296}{https://zenodo.org/records/17176296}. The code used to perform the simulations and generate all figures is available in the GitHub repository at \href{https://github.com/nereagurru/CCformation}{https://github.com/nereagurru/CCformation}.
\begin{acknowledgments}
The authors thank Jan Hellmann, Yves Marrocchi, Alessandro Morbidelli, and Fridolin Spitzer for valuable discussion. The authors also thank the anonymous referee for their comments that helped to improve the manuscript. J.D. and V.V. are funded by the European Union under the European Union’s Horizon Europe Research \& Innovation Programme 101040037 (PLANETOIDS). T. K. acknowledges the support from the ERC (project number 101019380- HolyEarth).

\end{acknowledgments}

\begin{contribution}

N.G.\;developed the model, performed the simulations, and led the writing of the
manuscript. J.D.\;and T.K.\;conceived the project. J.D.\;and V.V.\;contributed an earlier version of the code. T.K.\;provided expertise on the isotopic composition of meteorites. All authors contributed to interpreting the results and reviewing the manuscript.

\end{contribution}

\software{astropy \citep{astropy:2013, astropy:2018, astropy:2022},  
          numpy \citep{numpy}, 
          matplotlib \citep{matplotlib},
          DD-Diskevol \citep{Drazkowska2018},
          mcdust \citep{Vaikundaraman2025}.
          }
\bibliography{sample701}{}

@ARTICLE{Hellmann2023,
       author = {{Hellmann}, Jan L. and {Schneider}, Jonas M. and {W{\"o}lfer}, Elias and {Dr{\k{a}}{\.z}kowska}, Joanna and {Jansen}, Christian A. and {Hopp}, Timo and {Burkhardt}, Christoph and {Kleine}, Thorsten},
        title = "{Origin of Isotopic Diversity among Carbonaceous Chondrites}",
      journal = {\apjl},
         year = 2023,
        month = apr,
       volume = {946},
       number = {2},
          eid = {L34},
        pages = {L34},
          doi = {10.3847/2041-8213/acc102},
archivePrefix = {arXiv},
       eprint = {2303.04173},
 primaryClass = {astro-ph.EP},
       adsurl = {https://ui.adsabs.harvard.edu/abs/2023ApJ...946L..34H}
}

@ARTICLE{Desch18,
       author = {{Desch}, Steven J. and {Kalyaan}, Anusha and {O'D. Alexander}, Conel M.},
        title = "{The Effect of Jupiter's Formation on the Distribution of Refractory Elements and Inclusions in Meteorites}",
      journal = {\apjs},
         year = 2018,
        month = sep,
       volume = {238},
       number = {1},
          eid = {11},
        pages = {11},
          doi = {10.3847/1538-4365/aad95f},
archivePrefix = {arXiv},
       eprint = {1710.03809},
 primaryClass = {astro-ph.EP},
       adsurl = {https://ui.adsabs.harvard.edu/abs/2018ApJS..238...11D}
}

@ARTICLE{Drazkowska2013,
       author = {{Dr{\k{a}}{\.z}kowska}, J. and {Windmark}, F. and {Dullemond}, C.~P.},
        title = "{Planetesimal formation via sweep-up growth at the inner edge of dead zones}",
      journal = {\aap},
         year = 2013,
        month = aug,
       volume = {556},
          eid = {A37},
        pages = {A37},
          doi = {10.1051/0004-6361/201321566},
archivePrefix = {arXiv},
       eprint = {1306.3412},
 primaryClass = {astro-ph.EP},
       adsurl = {https://ui.adsabs.harvard.edu/abs/2013A&A...556A..37D}
}

@ARTICLE{Kleine2020,
       author = {{Kleine}, T. and {Budde}, G. and {Burkhardt}, C. and {Kruijer}, T.~S. and {Worsham}, E.~A. and {Morbidelli}, A. and {Nimmo}, F.},
        title = "{The Non-carbonaceous-Carbonaceous Meteorite Dichotomy}",
      journal = {\ssr},
         year = 2020,
        month = may,
       volume = {216},
       number = {4},
          eid = {55},
        pages = {55},
          doi = {10.1007/s11214-020-00675-w},
       adsurl = {https://ui.adsabs.harvard.edu/abs/2020SSRv..216...55K}
}

@ARTICLE{Shakura1973,
       author = {{Shakura}, N.~I. and {Sunyaev}, R.~A.},
        title = "{Black holes in binary systems. Observational appearance.}",
      journal = {\aap},
         year = 1973,
        month = jan,
       volume = {24},
        pages = {337-355},
       adsurl = {https://ui.adsabs.harvard.edu/abs/1973A&A....24..337S}
}

@ARTICLE{Hueso2005,
       author = {{Hueso}, R. and {Guillot}, T.},
        title = "{Evolution of protoplanetary disks: constraints from DM Tauri and GM Aurigae}",
      journal = {\aap},
         year = 2005,
        month = nov,
       volume = {442},
       number = {2},
        pages = {703-725},
          doi = {10.1051/0004-6361:20041905},
archivePrefix = {arXiv},
       eprint = {astro-ph/0506496},
 primaryClass = {astro-ph},
       adsurl = {https://ui.adsabs.harvard.edu/abs/2005A&A...442..703H}
}

@ARTICLE{Picogna2021,
       author = {{Picogna}, Giovanni and {Ercolano}, Barbara and {Espaillat}, Catherine C.},
        title = "{The dispersal of protoplanetary discs - III. Influence of stellar mass on disc photoevaporation}",
      journal = {\mnras},
         year = 2021,
        month = dec,
       volume = {508},
       number = {3},
        pages = {3611-3619},
          doi = {10.1093/mnras/stab2883},
archivePrefix = {arXiv},
       eprint = {2110.01250},
 primaryClass = {astro-ph.EP},
       adsurl = {https://ui.adsabs.harvard.edu/abs/2021MNRAS.508.3611P}
}

@ARTICLE{Jongejan2023,
       author = {{Jongejan}, S. and {Dominik}, C. and {Dullemond}, C.~P.},
        title = "{The effect of Jupiter on the CAI storage problem}",
      journal = {\aap},
         year = 2023,
        month = nov,
       volume = {679},
          eid = {A45},
        pages = {A45},
          doi = {10.1051/0004-6361/202245005},
archivePrefix = {arXiv},
       eprint = {2309.13760},
 primaryClass = {astro-ph.EP},
       adsurl = {https://ui.adsabs.harvard.edu/abs/2023A&A...679A..45J}
}

@ARTICLE{Dunham2023,
       author = {{Dunham}, E.~T. and {Sheikh}, A. and {Opara}, D. and {Matsuda}, N. and {Liu}, M. -C. and {McKeegan}, K.~D.},
        title = "{Calcium-aluminum-rich inclusions in non-carbonaceous chondrites: Abundances, sizes, and mineralogy}",
      journal = {\maps},
         year = 2023,
        month = may,
       volume = {58},
       number = {5},
        pages = {643-671},
          doi = {10.1111/maps.13975},
       adsurl = {https://ui.adsabs.harvard.edu/abs/2023M&PS...58..643D}
}

@ARTICLE{Stammler2023,
       author = {{Stammler}, Sebastian Markus and {Lichtenberg}, Tim and {Dr{\k{a}}{\.z}kowska}, Joanna and {Birnstiel}, Tilman},
        title = "{Leaky dust traps: How fragmentation impacts dust filtering by planets}",
      journal = {\aap},
         year = 2023,
        month = feb,
       volume = {670},
          eid = {L5},
        pages = {L5},
          doi = {10.1051/0004-6361/202245512},
archivePrefix = {arXiv},
       eprint = {2301.05505},
 primaryClass = {astro-ph.EP},
       adsurl = {https://ui.adsabs.harvard.edu/abs/2023A&A...670L...5S}
}

@ARTICLE{Haugbolle2019,
       author = {{Haugb{\o}lle}, Troels and {Weber}, Philipp and {Wielandt}, Daniel P. and {Ben{\'\i}tez-Llambay}, Pablo and {Bizzarro}, Martin and {Gressel}, Oliver and {Pessah}, Martin E.},
        title = "{Probing the Protosolar Disk Using Dust Filtering at Gaps in the Early Solar System}",
      journal = {\aj},
         year = 2019,
        month = aug,
       volume = {158},
       number = {2},
          eid = {55},
        pages = {55},
          doi = {10.3847/1538-3881/ab1591},
archivePrefix = {arXiv},
       eprint = {1903.12274},
 primaryClass = {astro-ph.EP},
       adsurl = {https://ui.adsabs.harvard.edu/abs/2019AJ....158...55H}
}

@ARTICLE{Weber2018,
       author = {{Weber}, Philipp and {Ben{\'\i}tez-Llambay}, Pablo and {Gressel}, Oliver and {Krapp}, Leonardo and {Pessah}, Martin E.},
        title = "{Characterizing the Variable Dust Permeability of Planet-induced Gaps}",
      journal = {\apj},
         year = 2018,
        month = feb,
       volume = {854},
       number = {2},
          eid = {153},
        pages = {153},
          doi = {10.3847/1538-4357/aaab63},
archivePrefix = {arXiv},
       eprint = {1801.07971},
 primaryClass = {astro-ph.EP},
       adsurl = {https://ui.adsabs.harvard.edu/abs/2018ApJ...854..153W}
}

@ARTICLE{Clepper2025,
       author = {{Van Clepper}, Eric and {Price}, Ellen M. and {Ciesla}, Fred J.},
        title = "{Three-dimensional Transport of Solids in a Protoplanetary Disk Containing a Growing Giant Planet}",
      journal = {\apj},
         year = 2025,
        month = feb,
       volume = {980},
       number = {2},
          eid = {201},
        pages = {201},
          doi = {10.3847/1538-4357/ada8a4},
archivePrefix = {arXiv},
       eprint = {2501.07520},
 primaryClass = {astro-ph.EP},
       adsurl = {https://ui.adsabs.harvard.edu/abs/2025ApJ...980..201V}
}

@ARTICLE{Carrera2017,
       author = {{Carrera}, Daniel and {Gorti}, Uma and {Johansen}, Anders and {Davies}, Melvyn B.},
        title = "{Planetesimal Formation by the Streaming Instability in a Photoevaporating Disk}",
      journal = {\apj},
         year = 2017,
        month = apr,
       volume = {839},
       number = {1},
          eid = {16},
        pages = {16},
          doi = {10.3847/1538-4357/aa6932},
archivePrefix = {arXiv},
       eprint = {1703.07895},
 primaryClass = {astro-ph.EP},
       adsurl = {https://ui.adsabs.harvard.edu/abs/2017ApJ...839...16C}
}

@ARTICLE{Lau2024,
       author = {{Lau}, Tommy Chi Ho and {Birnstiel}, Til and {Dr{\k{a}}{\.z}kowska}, Joanna and {Stammler}, Sebastian Markus},
        title = "{Sequential giant planet formation initiated by disc substructure}",
      journal = {\aap},
         year = 2024,
        month = aug,
       volume = {688},
          eid = {A22},
        pages = {A22},
          doi = {10.1051/0004-6361/202450464},
archivePrefix = {arXiv},
       eprint = {2406.12340},
 primaryClass = {astro-ph.EP},
       adsurl = {https://ui.adsabs.harvard.edu/abs/2024A&A...688A..22L}
}

@ARTICLE{Zsom2008,
       author = {{Zsom}, A. and {Dullemond}, C.~P.},
        title = "{A representative particle approach to coagulation and fragmentation of dust aggregates and fluid droplets}",
      journal = {\aap},
         year = 2008,
        month = oct,
       volume = {489},
       number = {2},
        pages = {931-941},
          doi = {10.1051/0004-6361:200809921},
archivePrefix = {arXiv},
       eprint = {0807.5052},
 primaryClass = {astro-ph},
       adsurl = {https://ui.adsabs.harvard.edu/abs/2008A&A...489..931Z}
}

@ARTICLE{Birnstiel2024,
       author = {{Birnstiel}, Tilman},
        title = "{Dust Growth and Evolution in Protoplanetary Disks}",
      journal = {\araa},
         year = 2024,
        month = sep,
       volume = {62},
       number = {1},
        pages = {157-202},
          doi = {10.1146/annurev-astro-071221-052705},
archivePrefix = {arXiv},
       eprint = {2312.13287},
 primaryClass = {astro-ph.EP},
       adsurl = {https://ui.adsabs.harvard.edu/abs/2024ARA&A..62..157B}
}

@ARTICLE{Bai2010,
       author = {{Bai}, Xue-Ning and {Stone}, James M.},
        title = "{Dynamics of Solids in the Midplane of Protoplanetary Disks: Implications for Planetesimal Formation}",
      journal = {\apj},
         year = 2010,
        month = oct,
       volume = {722},
       number = {2},
        pages = {1437-1459},
          doi = {10.1088/0004-637X/722/2/1437},
archivePrefix = {arXiv},
       eprint = {1005.4982},
 primaryClass = {astro-ph.EP},
       adsurl = {https://ui.adsabs.harvard.edu/abs/2010ApJ...722.1437B}
}

@ARTICLE{Beitz2012,
       author = {{Beitz}, E. and {G{\"u}ttler}, C. and {Weidling}, R. and {Blum}, J.},
        title = "{Free collisions in a microgravity many-particle experiment - II: The collision dynamics of dust-coated chondrules}",
      journal = {\icarus},
         year = 2012,
        month = mar,
       volume = {218},
       number = {1},
        pages = {701-706},
          doi = {10.1016/j.icarus.2011.11.036},
archivePrefix = {arXiv},
       eprint = {1105.3897},
 primaryClass = {astro-ph.EP},
       adsurl = {https://ui.adsabs.harvard.edu/abs/2012Icar..218..701B}
}

@ARTICLE{Machii2013,
       author = {{Machii}, Nagisa and {Nakamura}, Akiko M. and {G{\"u}ttler}, Carsten and {Beger}, Dirk and {Blum}, J{\"u}rgen},
        title = "{Collision of a chondrule with matrix: Relation between static strength of matrix and impact pressure}",
      journal = {\icarus},
         year = 2013,
        month = sep,
       volume = {226},
       number = {1},
        pages = {111-118},
          doi = {10.1016/j.icarus.2013.05.006},
       adsurl = {https://ui.adsabs.harvard.edu/abs/2013Icar..226..111M}
}

@ARTICLE{Umstatter2021,
       author = {{Umst{\"a}tter}, Philipp and {Urbassek}, Herbert M.},
        title = "{Granular mechanics simulations of collisions between chondritic aggregates}",
      journal = {\aap},
         year = 2021,
        month = aug,
       volume = {652},
          eid = {A40},
        pages = {A40},
          doi = {10.1051/0004-6361/202141581},
       adsurl = {https://ui.adsabs.harvard.edu/abs/2021A&A...652A..40U}
}

@ARTICLE{Guttler2010,
       author = {{G{\"u}ttler}, C. and {Blum}, J. and {Zsom}, A. and {Ormel}, C.~W. and {Dullemond}, C.~P.},
        title = "{The outcome of protoplanetary dust growth: pebbles, boulders, or planetesimals?. I. Mapping the zoo of laboratory collision experiments}",
      journal = {\aap},
         year = 2010,
        month = apr,
       volume = {513},
          eid = {A56},
        pages = {A56},
          doi = {10.1051/0004-6361/200912852},
archivePrefix = {arXiv},
       eprint = {0910.4251},
 primaryClass = {astro-ph.EP},
       adsurl = {https://ui.adsabs.harvard.edu/abs/2010A&A...513A..56G}
}

@ARTICLE{Ansdell2016,
       author = {{Ansdell}, M. and {Williams}, J.~P. and {van der Marel}, N. and {Carpenter}, J.~M. and {Guidi}, G. and {Hogerheijde}, M. and {Mathews}, G.~S. and {Manara}, C.~F. and {Miotello}, A. and {Natta}, A. and {Oliveira}, I. and {Tazzari}, M. and {Testi}, L. and {van Dishoeck}, E.~F. and {van Terwisga}, S.~E.},
        title = "{ALMA Survey of Lupus Protoplanetary Disks. I. Dust and Gas Masses}",
      journal = {\apj},
         year = 2016,
        month = sep,
       volume = {828},
       number = {1},
          eid = {46},
        pages = {46},
          doi = {10.3847/0004-637X/828/1/46},
archivePrefix = {arXiv},
       eprint = {1604.05719},
 primaryClass = {astro-ph.EP},
       adsurl = {https://ui.adsabs.harvard.edu/abs/2016ApJ...828...46A}
}

@ARTICLE{Lim2024,
       author = {{Lim}, Jeonghoon and {Simon}, Jacob B. and {Li}, Rixin and {Armitage}, Philip J. and {Carrera}, Daniel and {Lyra}, Wladimir and {Rea}, David G. and {Yang}, Chao-Chin and {Youdin}, Andrew N.},
        title = "{Streaming Instability and Turbulence: Conditions for Planetesimal Formation}",
      journal = {\apj},
         year = 2024,
        month = jul,
       volume = {969},
       number = {2},
          eid = {130},
        pages = {130},
          doi = {10.3847/1538-4357/ad47a2},
archivePrefix = {arXiv},
       eprint = {2312.12508},
 primaryClass = {astro-ph.EP},
       adsurl = {https://ui.adsabs.harvard.edu/abs/2024ApJ...969..130L}
}

@ARTICLE{Drazkowska2016,
       author = {{Dr{\k{a}}{\.z}kowska}, J. and {Alibert}, Y. and {Moore}, B.},
        title = "{Close-in planetesimal formation by pile-up of drifting pebbles}",
      journal = {\aap},
         year = 2016,
        month = oct,
       volume = {594},
          eid = {A105},
        pages = {A105},
          doi = {10.1051/0004-6361/201628983},
archivePrefix = {arXiv},
       eprint = {1607.05734},
 primaryClass = {astro-ph.EP},
       adsurl = {https://ui.adsabs.harvard.edu/abs/2016A&A...594A.105D}
}

@ARTICLE{Simon2016,
       author = {{Simon}, Jacob B. and {Armitage}, Philip J. and {Li}, Rixin and {Youdin}, Andrew N.},
        title = "{The Mass and Size Distribution of Planetesimals Formed by the Streaming Instability. I. The Role of Self-gravity}",
      journal = {\apj},
         year = 2016,
        month = may,
       volume = {822},
       number = {1},
          eid = {55},
        pages = {55},
          doi = {10.3847/0004-637X/822/1/55},
archivePrefix = {arXiv},
       eprint = {1512.00009},
 primaryClass = {astro-ph.SR},
       adsurl = {https://ui.adsabs.harvard.edu/abs/2016ApJ...822...55S}
}

@ARTICLE{Cuzzi2003,
       author = {{Cuzzi}, Jeffrey N. and {Davis}, Sanford S. and {Dobrovolskis}, Anthony R.},
        title = "{Blowing in the wind. II. Creation and redistribution of refractory inclusions in a turbulent protoplanetary nebula}",
      journal = {\icarus},
         year = 2003,
        month = dec,
       volume = {166},
       number = {2},
        pages = {385-402},
          doi = {10.1016/j.icarus.2003.08.016},
       adsurl = {https://ui.adsabs.harvard.edu/abs/2003Icar..166..385C}
}

@ARTICLE{Ciesla2007,
       author = {{Ciesla}, Fred J.},
        title = "{Outward Transport of High-Temperature Materials Around the Midplane of the Solar Nebula}",
      journal = {Science},
         year = 2007,
        month = oct,
       volume = {318},
       number = {5850},
        pages = {613},
          doi = {10.1126/science.1147273},
       adsurl = {https://ui.adsabs.harvard.edu/abs/2007Sci...318..613C}
}

@ARTICLE{Bryson2021,
       author = {{Bryson}, James F.~J. and {Brennecka}, Gregory A.},
        title = "{Constraints on Chondrule Generation, Disk Dynamics, and Asteroid Accretion from the Compositions of Carbonaceous Meteorites}",
      journal = {\apj},
         year = 2021,
        month = may,
       volume = {912},
       number = {2},
          eid = {163},
        pages = {163},
          doi = {10.3847/1538-4357/abea12},
       adsurl = {https://ui.adsabs.harvard.edu/abs/2021ApJ...912..163B}
}

@ARTICLE{Marrocchi2022,
       author = {{Marrocchi}, Yves and {Piralla}, Maxime and {Regnault}, Maxence and {Batanova}, Valentina and {Villeneuve}, Johan and {Jacquet}, Emmanuel},
        title = "{Isotopic evidence for two chondrule generations in CR chondrites and their relationships to other carbonaceous chondrites}",
      journal = {Earth and Planetary Science Letters},
         year = 2022,
        month = sep,
       volume = {593},
          eid = {117683},
        pages = {117683},
          doi = {10.1016/j.epsl.2022.117683},
       adsurl = {https://ui.adsabs.harvard.edu/abs/2022E&PSL.59317683M}
}

@ARTICLE{Schneider2020,
       author = {{Schneider}, Jonas M. and {Burkhardt}, Christoph and {Marrocchi}, Yves and {Brennecka}, Gregory A. and {Kleine}, Thorsten},
        title = "{Early evolution of the solar accretion disk inferred from Cr-Ti-O isotopes in individual chondrules}",
      journal = {Earth and Planetary Science Letters},
         year = 2020,
        month = dec,
       volume = {551},
          eid = {116585},
        pages = {116585},
          doi = {10.1016/j.epsl.2020.116585},
archivePrefix = {arXiv},
       eprint = {2009.08684},
 primaryClass = {astro-ph.EP},
       adsurl = {https://ui.adsabs.harvard.edu/abs/2020E&PSL.55116585S}
}

@INCOLLECTION{Scott2014,
       author = {{Scott}, E.~R.~D. and {Krot}, A.~N.},
        title = "{Chondrites and Their Components}",
    booktitle = {Meteorites and Cosmochemical Processes},
         year = 2014,
       editor = {{Davis}, Andrew M.},
       volume = {1},
        pages = {65-137},
       adsurl = {https://ui.adsabs.harvard.edu/abs/2014mcp..book...65S}
}

@ARTICLE{Stammler2019,
       author = {{Stammler}, Sebastian M. and {Dr{\k{a}}{\.z}kowska}, Joanna and {Birnstiel}, Til and {Klahr}, Hubert and {Dullemond}, Cornelis P. and {Andrews}, Sean M.},
        title = "{The DSHARP Rings: Evidence of Ongoing Planetesimal Formation?}",
      journal = {\apjl},
         year = 2019,
        month = oct,
       volume = {884},
       number = {1},
          eid = {L5},
        pages = {L5},
          doi = {10.3847/2041-8213/ab4423},
archivePrefix = {arXiv},
       eprint = {1909.04674},
 primaryClass = {astro-ph.EP},
       adsurl = {https://ui.adsabs.harvard.edu/abs/2019ApJ...884L...5S}
}

@ARTICLE{Drazkowska2018,
       author = {{Dr{\k{a}}{\.z}kowska}, J. and {Dullemond}, C.~P.},
        title = "{Planetesimal formation during protoplanetary disk buildup}",
      journal = {\aap},
         year = 2018,
        month = jun,
       volume = {614},
          eid = {A62},
        pages = {A62},
          doi = {10.1051/0004-6361/201732221},
archivePrefix = {arXiv},
       eprint = {1803.00575},
 primaryClass = {astro-ph.EP},
       adsurl = {https://ui.adsabs.harvard.edu/abs/2018A&A...614A..62D}
}

@ARTICLE{Hartmann1998,
       author = {{Hartmann}, Lee and {Calvet}, Nuria and {Gullbring}, Erik and {D'Alessio}, Paola},
        title = "{Accretion and the Evolution of T Tauri Disks}",
      journal = {\apj},
         year = 1998,
        month = mar,
       volume = {495},
       number = {1},
        pages = {385-400},
          doi = {10.1086/305277},
       adsurl = {https://ui.adsabs.harvard.edu/abs/1998ApJ...495..385H}
}

@INPROCEEDINGS{Pinte2023,
       author = {{Pinte}, C. and {Teague}, R. and {Flaherty}, K. and {Hall}, C. and {Facchini}, S. and {Casassus}, S.},
        title = "{Kinematic Structures in Planet-Forming Disks}",
    booktitle = {Protostars and Planets VII},
         year = 2023,
       editor = {{Inutsuka}, S. and {Aikawa}, Y. and {Muto}, T. and {Tomida}, K. and {Tamura}, M.},
       series = {Astronomical Society of the Pacific Conference Series},
       volume = {534},
        month = jul,
        pages = {645},
          doi = {10.48550/arXiv.2203.09528},
archivePrefix = {arXiv},
       eprint = {2203.09528},
 primaryClass = {astro-ph.EP},
       adsurl = {https://ui.adsabs.harvard.edu/abs/2023ASPC..534..645P}
}

@ARTICLE{Hellmann2020,
       author = {{Hellmann}, Jan L. and {Hopp}, Timo and {Burkhardt}, Christoph and {Kleine}, Thorsten},
        title = "{Origin of volatile element depletion among carbonaceous chondrites}",
      journal = {Earth and Planetary Science Letters},
         year = 2020,
        month = nov,
       volume = {549},
          eid = {116508},
        pages = {116508},
          doi = {10.1016/j.epsl.2020.116508},
       adsurl = {https://ui.adsabs.harvard.edu/abs/2020E&PSL.54916508H}
}

@ARTICLE{Budde2018,
       author = {{Budde}, Gerrit and {Kruijer}, Thomas S. and {Kleine}, Thorsten},
        title = "{Hf-W chronology of CR chondrites: Implications for the timescales of chondrule formation and the distribution of $^{26}$Al in the solar nebula}",
      journal = {\gca},
         year = 2018,
        month = feb,
       volume = {222},
        pages = {284-304},
          doi = {10.1016/j.gca.2017.10.014},
       adsurl = {https://ui.adsabs.harvard.edu/abs/2018GeCoA.222..284B}
}

@ARTICLE{Schrader2017,
       author = {{Schrader}, Devin L. and {Nagashima}, Kazuhide and {Krot}, Alexander N. and {Ogliore}, Ryan C. and {Yin}, Qing-Zhu and {Amelin}, Yuri and {Stirling}, Claudine H. and {Kaltenbach}, Angela},
        title = "{Distribution of $^{26}$Al in the CR chondrite chondrule-forming region of the protoplanetary disk}",
      journal = {\gca},
         year = 2017,
        month = mar,
       volume = {201},
        pages = {275-302},
          doi = {10.1016/j.gca.2016.06.023},
       adsurl = {https://ui.adsabs.harvard.edu/abs/2017GeCoA.201..275S}
}

@ARTICLE{Houge2023,
       author = {{Houge}, Adrien and {Krijt}, Sebastiaan},
        title = "{Collisional evolution of dust and water ice in protoplanetary discs during and after an accretion outburst}",
      journal = {\mnras},
         year = 2023,
        month = jun,
       volume = {521},
       number = {4},
        pages = {5826-5845},
          doi = {10.1093/mnras/stad866},
archivePrefix = {arXiv},
       eprint = {2303.11318},
 primaryClass = {astro-ph.EP},
       adsurl = {https://ui.adsabs.harvard.edu/abs/2023MNRAS.521.5826H}
}

@ARTICLE{Dohnanyi1969,
       author = {{Dohnanyi}, J.~S.},
        title = "{Collisional Model of Asteroids and Their Debris}",
      journal = {\jgr},
         year = 1969,
        month = may,
       volume = {74},
        pages = {2531-2554},
          doi = {10.1029/JB074i010p02531},
       adsurl = {https://ui.adsabs.harvard.edu/abs/1969JGR....74.2531D}
}

@ARTICLE{Gunkelmann2017,
       author = {{Gunkelmann}, Nina and {Kataoka}, Akimasa and {Dullemond}, Cornelis P. and {Urbassek}, Herbert M.},
        title = "{Low-velocity collisions of chondrules: How a thin dust cover helps enhance the sticking probability}",
      journal = {\aap},
         year = 2017,
        month = mar,
       volume = {599},
          eid = {L4},
        pages = {L4},
          doi = {10.1051/0004-6361/201630155},
       adsurl = {https://ui.adsabs.harvard.edu/abs/2017A&A...599L...4G}
}

@ARTICLE{Eriksson2020,
       author = {{Eriksson}, Linn E.~J. and {Johansen}, Anders and {Liu}, Beibei},
        title = "{Pebble drift and planetesimal formation in protoplanetary discs with embedded planets}",
      journal = {\aap},
         year = 2020,
        month = mar,
       volume = {635},
          eid = {A110},
        pages = {A110},
          doi = {10.1051/0004-6361/201937037},
archivePrefix = {arXiv},
       eprint = {2001.11042},
 primaryClass = {astro-ph.EP},
       adsurl = {https://ui.adsabs.harvard.edu/abs/2020A&A...635A.110E}
}

@ARTICLE{Kruijer2017,
       author = {{Kruijer}, Thomas S. and {Burkhardt}, Christoph and {Budde}, Gerrit and {Kleine}, Thorsten},
        title = "{Age of Jupiter inferred from the distinct genetics and formation times of meteorites}",
      journal = {Proceedings of the National Academy of Science},
         year = 2017,
        month = jun,
       volume = {114},
       number = {26},
        pages = {6712-6716},
          doi = {10.1073/pnas.1704461114},
       adsurl = {https://ui.adsabs.harvard.edu/abs/2017PNAS..114.6712K}
}

@ARTICLE{Alexander2019,
       author = {{Alexander}, Conel M. O'D.},
        title = "{Quantitative models for the elemental and isotopic fractionations in chondrites: The carbonaceous chondrites}",
      journal = {\gca},
         year = 2019,
        month = jun,
       volume = {254},
        pages = {277-309},
          doi = {10.1016/j.gca.2019.02.008},
       adsurl = {https://ui.adsabs.harvard.edu/abs/2019GeCoA.254..277A}
}

@INPROCEEDINGS{Johansen2014,
       author = {{Johansen}, A. and {Blum}, J. and {Tanaka}, H. and {Ormel}, C. and {Bizzarro}, M. and {Rickman}, H.},
        title = "{The Multifaceted Planetesimal Formation Process}",
    booktitle = {Protostars and Planets VI},
         year = 2014,
       editor = {{Beuther}, Henrik and {Klessen}, Ralf S. and {Dullemond}, Cornelis P. and {Henning}, Thomas},
        month = jan,
        pages = {547-570},
          doi = {10.2458/azu_uapress_9780816531240-ch024},
archivePrefix = {arXiv},
       eprint = {1402.1344},
 primaryClass = {astro-ph.EP},
       adsurl = {https://ui.adsabs.harvard.edu/abs/2014prpl.conf..547J}
}

@ARTICLE{Jones2012,
       author = {{Jones}, Rhian H.},
        title = "{Petrographic constraints on the diversity of chondrule reservoirs in the protoplanetary disk}",
      journal = {\maps},
         year = 2012,
        month = jul,
       volume = {47},
       number = {7},
        pages = {1176-1190},
          doi = {10.1111/j.1945-5100.2011.01327.x},
       adsurl = {https://ui.adsabs.harvard.edu/abs/2012M&PS...47.1176J}
}

@ARTICLE{Birnstiel2012,
       author = {{Birnstiel}, T. and {Klahr}, H. and {Ercolano}, B.},
        title = "{A simple model for the evolution of the dust population in protoplanetary disks}",
      journal = {\aap},
         year = 2012,
        month = mar,
       volume = {539},
          eid = {A148},
        pages = {A148},
          doi = {10.1051/0004-6361/201118136},
archivePrefix = {arXiv},
       eprint = {1201.5781},
 primaryClass = {astro-ph.EP},
       adsurl = {https://ui.adsabs.harvard.edu/abs/2012A&A...539A.148B}
}

@ARTICLE{Morbidelli2007,
       author = {{Morbidelli}, Alessandro and {Crida}, Aur{\'e}lien},
        title = "{The dynamics of Jupiter and Saturn in the gaseous protoplanetary disk}",
      journal = {\icarus},
         year = 2007,
        month = nov,
       volume = {191},
       number = {1},
        pages = {158-171},
          doi = {10.1016/j.icarus.2007.04.001},
       adsurl = {https://ui.adsabs.harvard.edu/abs/2007Icar..191..158M}
}

@ARTICLE{Sugiura2014,
       author = {{Sugiura}, Naoji and {Fujiya}, Wataru},
        title = "{Correlated accretion ages and ɛ$^{54}$Cr of meteorite parent bodies and the evolution of the solar nebula}",
      journal = {\maps},
         year = 2014,
        month = may,
       volume = {49},
       number = {5},
        pages = {772-787},
          doi = {10.1111/maps.12292},
       adsurl = {https://ui.adsabs.harvard.edu/abs/2014M&PS...49..772S}
}

@ARTICLE{Fukuda2022,
       author = {{Fukuda}, Kohei and {Tenner}, Travis J. and {Kimura}, Makoto and {Tomioka}, Naotaka and {Siron}, Guillaume and {Ushikubo}, Takayuki and {Chaumard}, No{\"e}l and {Hertwig}, Andreas T. and {Kita}, Noriko T.},
        title = "{A temporal shift of chondrule generation from the inner to outer Solar System inferred from oxygen isotopes and Al-Mg chronology of chondrules from primitive CM and CO chondrites}",
      journal = {\gca},
         year = 2022,
        month = apr,
       volume = {322},
        pages = {194-226},
          doi = {10.1016/j.gca.2021.12.027},
       adsurl = {https://ui.adsabs.harvard.edu/abs/2022GeCoA.322..194F}
}

@ARTICLE{Pinilla2012,
       author = {{Pinilla}, P. and {Benisty}, M. and {Birnstiel}, T.},
        title = "{Ring shaped dust accumulation in transition disks}",
      journal = {\aap},
         year = 2012,
        month = sep,
       volume = {545},
          eid = {A81},
        pages = {A81},
          doi = {10.1051/0004-6361/201219315},
archivePrefix = {arXiv},
       eprint = {1207.6485},
 primaryClass = {astro-ph.EP},
       adsurl = {https://ui.adsabs.harvard.edu/abs/2012A&A...545A..81P}
}

@ARTICLE{Andrews2018,
       author = {{Andrews}, Sean M. and {Huang}, Jane and {P{\'e}rez}, Laura M. and {Isella}, Andrea and {Dullemond}, Cornelis P. and {Kurtovic}, Nicol{\'a}s T. and {Guzm{\'a}n}, Viviana V. and {Carpenter}, John M. and {Wilner}, David J. and {Zhang}, Shangjia and {Zhu}, Zhaohuan and {Birnstiel}, Tilman and {Bai}, Xue-Ning and {Benisty}, Myriam and {Hughes}, A. Meredith and {{\"O}berg}, Karin I. and {Ricci}, Luca},
        title = "{The Disk Substructures at High Angular Resolution Project (DSHARP). I. Motivation, Sample, Calibration, and Overview}",
      journal = {\apjl},
         year = 2018,
        month = dec,
       volume = {869},
       number = {2},
          eid = {L41},
        pages = {L41},
          doi = {10.3847/2041-8213/aaf741},
archivePrefix = {arXiv},
       eprint = {1812.04040},
 primaryClass = {astro-ph.SR},
       adsurl = {https://ui.adsabs.harvard.edu/abs/2018ApJ...869L..41A}
}

@ARTICLE{Lega2025,
       author = {{Lega}, E. and {Morbidelli}, A. and {Masset}, F. and {B{\'e}thune}, W.},
        title = "{On the formation of multiple dust-trapping rings in the inner Solar system}",
      journal = {arXiv e-prints},
         year = 2025,
        month = aug,
          eid = {arXiv:2508.02410},
        pages = {arXiv:2508.02410},
          doi = {10.48550/arXiv.2508.02410},
archivePrefix = {arXiv},
       eprint = {2508.02410},
 primaryClass = {astro-ph.EP},
       adsurl = {https://ui.adsabs.harvard.edu/abs/2025arXiv250802410L}
}

@ARTICLE{Vaikundaraman2025,
       author = {{Vaikundaraman}, Vignesh and {Gurrutxaga}, Nerea and {Dr{\k{a}}{\.z}kowska}, Joanna},
        title = "{mcdust: A 2D Monte Carlo code for dust coagulation in protoplanetary disks}",
      journal = {arXiv e-prints},
         year = 2025,
        month = jul,
          eid = {arXiv:2507.21239},
        pages = {arXiv:2507.21239},
          doi = {10.48550/arXiv.2507.21239},
archivePrefix = {arXiv},
       eprint = {2507.21239},
 primaryClass = {astro-ph.IM},
       adsurl = {https://ui.adsabs.harvard.edu/abs/2025arXiv250721239V}
}

@ARTICLE{Krijt2016,
       author = {{Krijt}, Sebastiaan and {Ciesla}, Fred J. and {Bergin}, Edwin A.},
        title = "{Tracing Water Vapor and Ice During Dust Growth}",
      journal = {\apj},
         year = 2016,
        month = dec,
       volume = {833},
       number = {2},
          eid = {285},
        pages = {285},
          doi = {10.3847/1538-4357/833/2/285},
archivePrefix = {arXiv},
       eprint = {1610.06463},
 primaryClass = {astro-ph.EP},
       adsurl = {https://ui.adsabs.harvard.edu/abs/2016ApJ...833..285K}
}

@INPROCEEDINGS{Drazkowska2023,
       author = {{Dr{\k{a}}{\.z}kowska}, J. and {Bitsch}, B. and {Lambrechts}, M. and {Mulders}, G.~D. and {Harsono}, D. and {Vazan}, A. and {Liu}, B. and {Ormel}, C.~W. and {Kretke}, K. and {Morbidelli}, A.},
        title = "{Planet Formation Theory in the Era of ALMA and Kepler: from Pebbles to Exoplanets}",
    booktitle = {Protostars and Planets VII},
         year = 2023,
       editor = {{Inutsuka}, S. and {Aikawa}, Y. and {Muto}, T. and {Tomida}, K. and {Tamura}, M.},
       series = {Astronomical Society of the Pacific Conference Series},
       volume = {534},
        month = jul,
        pages = {717},
          doi = {10.48550/arXiv.2203.09759},
archivePrefix = {arXiv},
       eprint = {2203.09759},
 primaryClass = {astro-ph.EP},
       adsurl = {https://ui.adsabs.harvard.edu/abs/2023ASPC..534..717D}
}

@ARTICLE{Youdin2005,
       author = {{Youdin}, Andrew N. and {Goodman}, Jeremy},
        title = "{Streaming Instabilities in Protoplanetary Disks}",
      journal = {\apj},
         year = 2005,
        month = feb,
       volume = {620},
       number = {1},
        pages = {459-469},
          doi = {10.1086/426895},
archivePrefix = {arXiv},
       eprint = {astro-ph/0409263},
 primaryClass = {astro-ph},
       adsurl = {https://ui.adsabs.harvard.edu/abs/2005ApJ...620..459Y}
}

@ARTICLE{Hopp2022,
       author = {{Hopp}, Timo and {Dauphas}, Nicolas and {Abe}, Yoshinari and {Al{\'e}on}, J{\'e}r{\^o}me and {O'D. Alexander}, Conel M. and {Amari}, Sachiko and {Amelin}, Yuri and {Bajo}, Ken-ichi and {Bizzarro}, Martin and {Bouvier}, Audrey and {Carlson}, Richard W. and {Chaussidon}, Marc and {Choi}, Byeon-Gak and {Davis}, Andrew M. and {Di Rocco}, Tommaso and {Fujiya}, Wataru and {Fukai}, Ryota and {Gautam}, Ikshu and {Haba}, Makiko K. and {Hibiya}, Yuki and {Hidaka}, Hiroshi and {Homma}, Hisashi and {Hoppe}, Peter and {Huss}, Gary R. and {Ichida}, Kiyohiro and {Iizuka}, Tsuyoshi and {Ireland}, Trevor R. and {Ishikawa}, Akira and {Ito}, Motoo and {Itoh}, Shoichi and {Kawasaki}, Noriyuki and {Kita}, Noriko T. and {Kitajima}, Kouki and {Kleine}, Thorsten and {Komatani}, Shintaro and {Krot}, Alexander N. and {Liu}, Ming-Chang and {Masuda}, Yuki and {McKeegan}, Kevin D. and {Morita}, Mayu and {Motomura}, Kazuko and {Moynier}, Fr{\'e}d{\'e}ric and {Nakai}, Izumi and {Nagashima}, Kazuhide and {Nesvorn{\'y}}, David and {Nguyen}, Ann and {Nittler}, Larry and {Onose}, Morihiko and {Pack}, Andreas and {Park}, Changkun and {Piani}, Laurette and {Qin}, Liping and {Russell}, Sara S. and {Sakamoto}, Naoya and {Sch{\"o}nb{\"a}chler}, Maria and {Tafla}, Lauren and {Tang}, Haolan and {Terada}, Kentaro and {Terada}, Yasuko and {Usui}, Tomohiro and {Wada}, Sohei and {Wadhwa}, Meenakshi and {Walker}, Richard J. and {Yamashita}, Katsuyuki and {Yin}, Qing-Zhu and {Yokoyama}, Tetsuya and {Yoneda}, Shigekazu and {Young}, Edward D. and {Yui}, Hiroharu and {Zhang}, Ai-Cheng and {Nakamura}, Tomoki and {Naraoka}, Hiroshi and {Noguchi}, Takaaki and {Okazaki}, Ryuji and {Sakamoto}, Kanako and {Yabuta}, Hikaru and {Abe}, Masanao and {Miyazaki}, Akiko and {Nakato}, Aiko and {Nishimura}, Masahiro and {Okada}, Tatsuaki and {Yada}, Toru and {Yogata}, Kasumi and {Nakazawa}, Satoru and {Saiki}, Takanao and {Tanaka}, Satoshi and {Terui}, Fuyuto and {Tsuda}, Yuichi and {Watanabe}, Sei-ichiro and {Yoshikawa}, Makoto and {Tachibana}, Shogo and {Yurimoto}, Hisayoshi},
        title = "{Ryugu's nucleosynthetic heritage from the outskirts of the Solar System}",
      journal = {Science Advances},
         year = 2022,
        month = nov,
       volume = {8},
       number = {46},
          eid = {eadd8141},
        pages = {eadd8141},
          doi = {10.1126/sciadv.add8141},
       adsurl = {https://ui.adsabs.harvard.edu/abs/2022SciA....8D8141H}
}

@ARTICLE{Spitzer2025,
       author = {{Spitzer}, Fridolin and {Hopp}, Timo and {Burkhardt}, Christoph and {Dauphas}, Nicolas and {Kleine}, Thorsten},
        title = "{The evolution of planetesimal reservoirs revealed by Fe-Ni isotope anomalies in differentiated meteorites}",
      journal = {Earth and Planetary Science Letters},
         year = 2025,
        month = oct,
       volume = {667},
          eid = {119530},
        pages = {119530},
          doi = {10.1016/j.epsl.2025.119530},
       adsurl = {https://ui.adsabs.harvard.edu/abs/2025E&PSL.66719530S}
}

@ARTICLE{Ercolano2017,
       author = {{Ercolano}, Barbara and {Jennings}, Jeff and {Rosotti}, Giovanni and {Birnstiel}, Tilman},
        title = "{X-ray photoevaporation's limited success in the formation of planetesimals by the streaming instability}",
      journal = {\mnras},
         year = 2017,
        month = dec,
       volume = {472},
       number = {4},
        pages = {4117-4125},
          doi = {10.1093/mnras/stx2294},
archivePrefix = {arXiv},
       eprint = {1709.00361},
 primaryClass = {astro-ph.EP},
       adsurl = {https://ui.adsabs.harvard.edu/abs/2017MNRAS.472.4117E}
}

@ARTICLE{Kleine2005,
       author = {{Kleine}, Thorsten and {Mezger}, Klaus and {Palme}, Herbert and {Scherer}, Erik and {M{\"u}nker}, Carsten},
        title = "{Early core formation in asteroids and late accretion of chondrite parent bodies: Evidence from $^{182}$Hf- $^{182}$W in CAIs, metal-rich chondrites, and iron meteorites}",
      journal = {\gca},
         year = 2005,
        month = dec,
       volume = {69},
       number = {24},
        pages = {5805-5818},
          doi = {10.1016/j.gca.2005.07.012},
       adsurl = {https://ui.adsabs.harvard.edu/abs/2005GeCoA..69.5805K}
}

@ARTICLE{Kruijer2014,
       author = {{Kruijer}, T.~S. and {Touboul}, M. and {Fischer-G{\"o}dde}, M. and {Bermingham}, K.~R. and {Walker}, R.~J. and {Kleine}, T.},
        title = "{Protracted core formation and rapid accretion of protoplanets}",
      journal = {Science},
         year = 2014,
        month = jun,
       volume = {344},
       number = {6188},
        pages = {1150-1154},
          doi = {10.1126/science.1251766},
       adsurl = {https://ui.adsabs.harvard.edu/abs/2014Sci...344.1150K}
}

@article{Spitzer2024,
author = {Fridolin Spitzer  and Thorsten Kleine  and Christoph Burkhardt  and Timo Hopp  and Tetsuya Yokoyama  and Yoshinari Abe  and Jérôme Aléon  and Conel M. O’D Alexander  and Sachiko Amari  and Yuri Amelin  and Ken-ichi Bajo  and Martin Bizzarro  and Audrey Bouvier  and Richard W. Carlson  and Marc Chaussidon  and Byeon-Gak Choi  and Nicolas Dauphas  and Andrew M. Davis  and Tommaso Di Rocco  and Wataru Fujiya  and Ryota Fukai  and Ikshu Gautam  and Makiko K. Haba  and Yuki Hibiya  and Hiroshi Hidaka  and Hisashi Homma  and Peter Hoppe  and Gary R. Huss  and Kiyohiro Ichida  and Tsuyoshi Iizuka  and Trevor R. Ireland  and Akira Ishikawa  and Shoichi Itoh  and Noriyuki Kawasaki  and Noriko T. Kita  and Kouki Kitajima  and Shintaro Komatani  and Alexander N. Krot  and Ming-Chang Liu  and Yuki Masuda  and Mayu Morita  and Fréderic Moynier  and Kazuko Motomura  and Izumi Nakai  and Kazuhide Nagashima  and Ann Nguyen  and Larry Nittler  and Morihiko Onose  and Andreas Pack  and Changkun Park  and Laurette Piani  and Liping Qin  and Sara S. Russell  and Naoya Sakamoto  and Maria Schönbächler  and Lauren Tafla  and Haolan Tang  and Kentaro Terada  and Yasuko Terada  and Tomohiro Usui  and Sohei Wada  and Meenakshi Wadhwa  and Richard J. Walker  and Katsuyuki Yamashita  and Qing-Zhu Yin  and Shigekazu Yoneda  and Edward D. Young  and Hiroharu Yui  and Ai-Cheng Zhang  and Tomoki Nakamura  and Hiroshi Naraoka  and Takaaki Noguchi  and Ryuji Okazaki  and Kanako Sakamoto  and Hikaru Yabuta  and Masanao Abe  and Akiko Miyazaki  and Aiko Nakato  and Masahiro Nishimura  and Tatsuaki Okada  and Toru Yada  and Kasumi Yogata  and Satoru Nakazawa  and Takanao Saiki  and Satoshi Tanaka  and Fuyuto Terui  and Yuichi Tsuda  and Sei-ichiro Watanabe  and Makoto Yoshikawa  and Shogo Tachibana  and Hisayoshi Yurimoto },
title = {The Ni isotopic composition of Ryugu reveals a common accretion region for carbonaceous chondrites},
journal = {Science Advances},
volume = {10},
number = {39},
pages = {eadp2426},
year = {2024},
doi = {10.1126/sciadv.adp2426}}

@ARTICLE{Pfeil2025,
       author = {{Pfeil}, Thomas and {Armitage}, Philip J. and {Jiang}, Yan-Fei},
        title = "{Fragmentation-limited Dust Filtration in 2D Simulations of Planet─Disk Systems with Dust Coagulation: Parameter Study and Implications for the Inner Disk's Dust Mass Budget and Composition}",
      journal = {\apj},
         year = 2025,
        month = dec,
       volume = {994},
       number = {2},
          eid = {272},
        pages = {272},
          doi = {10.3847/1538-4357/ae1295},
archivePrefix = {arXiv},
       eprint = {2510.08574},
 primaryClass = {astro-ph.EP},
       adsurl = {https://ui.adsabs.harvard.edu/abs/2025ApJ...994..272P}
}

@ARTICLE{Weiss2021,
       author = {{Weiss}, Benjamin P. and {Bai}, Xue-Ning and {Fu}, Roger R.},
        title = "{History of the solar nebula from meteorite paleomagnetism}",
      journal = {Science Advances},
         year = 2021,
        month = jan,
       volume = {7},
       number = {1},
        pages = {eaba5967},
          doi = {10.1126/sciadv.aba5967},
archivePrefix = {arXiv},
       eprint = {2103.02011},
 primaryClass = {astro-ph.EP},
       adsurl = {https://ui.adsabs.harvard.edu/abs/2021SciA....7.5967W}
}

@ARTICLE{Ormel2008,
       author = {{Ormel}, C.~W. and {Cuzzi}, J.~N. and {Tielens}, A.~G.~G.~M.},
        title = "{Co-Accretion of Chondrules and Dust in the Solar Nebula}",
      journal = {\apj},
         year = 2008,
        month = jun,
       volume = {679},
       number = {2},
        pages = {1588-1610},
          doi = {10.1086/587836},
archivePrefix = {arXiv},
       eprint = {0802.4048},
 primaryClass = {astro-ph},
       adsurl = {https://ui.adsabs.harvard.edu/abs/2008ApJ...679.1588O}
}

@article{astropy:2013,
Adsurl = {http://adsabs.harvard.edu/abs/2013A%26A...558A..33A},
Archiveprefix = {arXiv},
Author = {{Astropy Collaboration} and {Robitaille}, T.~P. and {Tollerud}, E.~J. and {Greenfield}, P. and {Droettboom}, M. and {Bray}, E. and {Aldcroft}, T. and {Davis}, M. and {Ginsburg}, A. and {Price-Whelan}, A.~M. and {Kerzendorf}, W.~E. and {Conley}, A. and {Crighton}, N. and {Barbary}, K. and {Muna}, D. and {Ferguson}, H. and {Grollier}, F. and {Parikh}, M.~M. and {Nair}, P.~H. and {Unther}, H.~M. and {Deil}, C. and {Woillez}, J. and {Conseil}, S. and {Kramer}, R. and {Turner}, J.~E.~H. and {Singer}, L. and {Fox}, R. and {Weaver}, B.~A. and {Zabalza}, V. and {Edwards}, Z.~I. and {Azalee Bostroem}, K. and {Burke}, D.~J. and {Casey}, A.~R. and {Crawford}, S.~M. and {Dencheva}, N. and {Ely}, J. and {Jenness}, T. and {Labrie}, K. and {Lim}, P.~L. and {Pierfederici}, F. and {Pontzen}, A. and {Ptak}, A. and {Refsdal}, B. and {Servillat}, M. and {Streicher}, O.},
Doi = {10.1051/0004-6361/201322068},
Eid = {A33},
Eprint = {1307.6212},
Journal = {\aap},
Month = oct,
Pages = {A33},
Primaryclass = {astro-ph.IM},
Title = {{Astropy: A community Python package for astronomy}},
Volume = 558,
Year = 2013}

@ARTICLE{astropy:2018,
       author = {{Astropy Collaboration} and {Price-Whelan}, A.~M. and
         {Sip{\H{o}}cz}, B.~M. and {G{\"u}nther}, H.~M. and {Lim}, P.~L. and
         {Crawford}, S.~M. and {Conseil}, S. and {Shupe}, D.~L. and
         {Craig}, M.~W. and {Dencheva}, N. and {Ginsburg}, A. and {Vand
        erPlas}, J.~T. and {Bradley}, L.~D. and {P{\'e}rez-Su{\'a}rez}, D. and
         {de Val-Borro}, M. and {Aldcroft}, T.~L. and {Cruz}, K.~L. and
         {Robitaille}, T.~P. and {Tollerud}, E.~J. and {Ardelean}, C. and
         {Babej}, T. and {Bach}, Y.~P. and {Bachetti}, M. and {Bakanov}, A.~V. and
         {Bamford}, S.~P. and {Barentsen}, G. and {Barmby}, P. and
         {Baumbach}, A. and {Berry}, K.~L. and {Biscani}, F. and {Boquien}, M. and
         {Bostroem}, K.~A. and {Bouma}, L.~G. and {Brammer}, G.~B. and
         {Bray}, E.~M. and {Breytenbach}, H. and {Buddelmeijer}, H. and
         {Burke}, D.~J. and {Calderone}, G. and {Cano Rodr{\'\i}guez}, J.~L. and
         {Cara}, M. and {Cardoso}, J.~V.~M. and {Cheedella}, S. and {Copin}, Y. and
         {Corrales}, L. and {Crichton}, D. and {D'Avella}, D. and {Deil}, C. and
         {Depagne}, {\'E}. and {Dietrich}, J.~P. and {Donath}, A. and
         {Droettboom}, M. and {Earl}, N. and {Erben}, T. and {Fabbro}, S. and
         {Ferreira}, L.~A. and {Finethy}, T. and {Fox}, R.~T. and
         {Garrison}, L.~H. and {Gibbons}, S.~L.~J. and {Goldstein}, D.~A. and
         {Gommers}, R. and {Greco}, J.~P. and {Greenfield}, P. and
         {Groener}, A.~M. and {Grollier}, F. and {Hagen}, A. and {Hirst}, P. and
         {Homeier}, D. and {Horton}, A.~J. and {Hosseinzadeh}, G. and {Hu}, L. and
         {Hunkeler}, J.~S. and {Ivezi{\'c}}, {\v{Z}}. and {Jain}, A. and
         {Jenness}, T. and {Kanarek}, G. and {Kendrew}, S. and {Kern}, N.~S. and
         {Kerzendorf}, W.~E. and {Khvalko}, A. and {King}, J. and {Kirkby}, D. and
         {Kulkarni}, A.~M. and {Kumar}, A. and {Lee}, A. and {Lenz}, D. and
         {Littlefair}, S.~P. and {Ma}, Z. and {Macleod}, D.~M. and
         {Mastropietro}, M. and {McCully}, C. and {Montagnac}, S. and
         {Morris}, B.~M. and {Mueller}, M. and {Mumford}, S.~J. and {Muna}, D. and
         {Murphy}, N.~A. and {Nelson}, S. and {Nguyen}, G.~H. and
         {Ninan}, J.~P. and {N{\"o}the}, M. and {Ogaz}, S. and {Oh}, S. and
         {Parejko}, J.~K. and {Parley}, N. and {Pascual}, S. and {Patil}, R. and
         {Patil}, A.~A. and {Plunkett}, A.~L. and {Prochaska}, J.~X. and
         {Rastogi}, T. and {Reddy Janga}, V. and {Sabater}, J. and
         {Sakurikar}, P. and {Seifert}, M. and {Sherbert}, L.~E. and
         {Sherwood-Taylor}, H. and {Shih}, A.~Y. and {Sick}, J. and
         {Silbiger}, M.~T. and {Singanamalla}, S. and {Singer}, L.~P. and
         {Sladen}, P.~H. and {Sooley}, K.~A. and {Sornarajah}, S. and
         {Streicher}, O. and {Teuben}, P. and {Thomas}, S.~W. and
         {Tremblay}, G.~R. and {Turner}, J.~E.~H. and {Terr{\'o}n}, V. and
         {van Kerkwijk}, M.~H. and {de la Vega}, A. and {Watkins}, L.~L. and
         {Weaver}, B.~A. and {Whitmore}, J.~B. and {Woillez}, J. and
         {Zabalza}, V. and {Astropy Contributors}},
        title = "{The Astropy Project: Building an Open-science Project and Status of the v2.0 Core Package}",
      journal = {\aj},
         year = 2018,
        month = sep,
       volume = {156},
       number = {3},
          eid = {123},
        pages = {123},
          doi = {10.3847/1538-3881/aabc4f},
archivePrefix = {arXiv},
       eprint = {1801.02634},
 primaryClass = {astro-ph.IM},
       adsurl = {https://ui.adsabs.harvard.edu/abs/2018AJ....156..123A}
}

@ARTICLE{astropy:2022,
       author = {{Astropy Collaboration} and {Price-Whelan}, Adrian M. and {Lim}, Pey Lian and {Earl}, Nicholas and {Starkman}, Nathaniel and {Bradley}, Larry and {Shupe}, David L. and {Patil}, Aarya A. and {Corrales}, Lia and {Brasseur}, C.~E. and {N{"o}the}, Maximilian and {Donath}, Axel and {Tollerud}, Erik and {Morris}, Brett M. and {Ginsburg}, Adam and {Vaher}, Eero and {Weaver}, Benjamin A. and {Tocknell}, James and {Jamieson}, William and {van Kerkwijk}, Marten H. and {Robitaille}, Thomas P. and {Merry}, Bruce and {Bachetti}, Matteo and {G{"u}nther}, H. Moritz and {Aldcroft}, Thomas L. and {Alvarado-Montes}, Jaime A. and {Archibald}, Anne M. and {B{'o}di}, Attila and {Bapat}, Shreyas and {Barentsen}, Geert and {Baz{'a}n}, Juanjo and {Biswas}, Manish and {Boquien}, M{'e}d{'e}ric and {Burke}, D.~J. and {Cara}, Daria and {Cara}, Mihai and {Conroy}, Kyle E. and {Conseil}, Simon and {Craig}, Matthew W. and {Cross}, Robert M. and {Cruz}, Kelle L. and {D'Eugenio}, Francesco and {Dencheva}, Nadia and {Devillepoix}, Hadrien A.~R. and {Dietrich}, J{"o}rg P. and {Eigenbrot}, Arthur Davis and {Erben}, Thomas and {Ferreira}, Leonardo and {Foreman-Mackey}, Daniel and {Fox}, Ryan and {Freij}, Nabil and {Garg}, Suyog and {Geda}, Robel and {Glattly}, Lauren and {Gondhalekar}, Yash and {Gordon}, Karl D. and {Grant}, David and {Greenfield}, Perry and {Groener}, Austen M. and {Guest}, Steve and {Gurovich}, Sebastian and {Handberg}, Rasmus and {Hart}, Akeem and {Hatfield-Dodds}, Zac and {Homeier}, Derek and {Hosseinzadeh}, Griffin and {Jenness}, Tim and {Jones}, Craig K. and {Joseph}, Prajwel and {Kalmbach}, J. Bryce and {Karamehmetoglu}, Emir and {Ka{l}uszy{'n}ski}, Miko{l}aj and {Kelley}, Michael S.~P. and {Kern}, Nicholas and {Kerzendorf}, Wolfgang E. and {Koch}, Eric W. and {Kulumani}, Shankar and {Lee}, Antony and {Ly}, Chun and {Ma}, Zhiyuan and {MacBride}, Conor and {Maljaars}, Jakob M. and {Muna}, Demitri and {Murphy}, N.~A. and {Norman}, Henrik and {O'Steen}, Richard and {Oman}, Kyle A. and {Pacifici}, Camilla and {Pascual}, Sergio and {Pascual-Granado}, J. and {Patil}, Rohit R. and {Perren}, Gabriel I. and {Pickering}, Timothy E. and {Rastogi}, Tanuj and {Roulston}, Benjamin R. and {Ryan}, Daniel F. and {Rykoff}, Eli S. and {Sabater}, Jose and {Sakurikar}, Parikshit and {Salgado}, Jes{'u}s and {Sanghi}, Aniket and {Saunders}, Nicholas and {Savchenko}, Volodymyr and {Schwardt}, Ludwig and {Seifert-Eckert}, Michael and {Shih}, Albert Y. and {Jain}, Anany Shrey and {Shukla}, Gyanendra and {Sick}, Jonathan and {Simpson}, Chris and {Singanamalla}, Sudheesh and {Singer}, Leo P. and {Singhal}, Jaladh and {Sinha}, Manodeep and {Sip{H{o}}cz}, Brigitta M. and {Spitler}, Lee R. and {Stansby}, David and {Streicher}, Ole and {{{S}}umak}, Jani and {Swinbank}, John D. and {Taranu}, Dan S. and {Tewary}, Nikita and {Tremblay}, Grant R. and {Val-Borro}, Miguel de and {Van Kooten}, Samuel J. and {Vasovi{'c}}, Zlatan and {Verma}, Shresth and {de Miranda Cardoso}, Jos{'e} Vin{'i}cius and {Williams}, Peter K.~G. and {Wilson}, Tom J. and {Winkel}, Benjamin and {Wood-Vasey}, W.~M. and {Xue}, Rui and {Yoachim}, Peter and {Zhang}, Chen and {Zonca}, Andrea and {Astropy Project Contributors}},
        title = "{The Astropy Project: Sustaining and Growing a Community-oriented Open-source Project and the Latest Major Release (v5.0) of the Core Package}",
      journal = {\apj},
         year = 2022,
        month = aug,
       volume = {935},
       number = {2},
          eid = {167},
        pages = {167},
          doi = {10.3847/1538-4357/ac7c74},
archivePrefix = {arXiv},
       eprint = {2206.14220},
 primaryClass = {astro-ph.IM},
       adsurl = {https://ui.adsabs.harvard.edu/abs/2022ApJ...935..167A}
}

@Article{         numpy,
 title         = {Array programming with {NumPy}},
 author        = {Charles R. Harris and K. Jarrod Millman and St{\'{e}}fan J.
                 van der Walt and Ralf Gommers and Pauli Virtanen and David
                 Cournapeau and Eric Wieser and Julian Taylor and Sebastian
                 Berg and Nathaniel J. Smith and Robert Kern and Matti Picus
                 and Stephan Hoyer and Marten H. van Kerkwijk and Matthew
                 Brett and Allan Haldane and Jaime Fern{\'{a}}ndez del
                 R{\'{i}}o and Mark Wiebe and Pearu Peterson and Pierre
                 G{\'{e}}rard-Marchant and Kevin Sheppard and Tyler Reddy and
                 Warren Weckesser and Hameer Abbasi and Christoph Gohlke and
                 Travis E. Oliphant},
 year          = {2020},
 month         = sep,
 journal       = {Nature},
 volume        = {585},
 number        = {7825},
 pages         = {357--362},
 doi           = {10.1038/s41586-020-2649-2},
 publisher     = {Springer Science and Business Media {LLC}},
 url           = {https://doi.org/10.1038/s41586-020-2649-2}
}

@Article{matplotlib,
  Author    = {Hunter, J. D.},
  Title     = {Matplotlib: A 2D graphics environment},
  Journal   = {Computing in Science \& Engineering},
  Volume    = {9},
  Number    = {3},
  Pages     = {90--95},
  publisher = {IEEE COMPUTER SOC},
  doi       = {10.1109/MCSE.2007.55},
  year      = 2007
}

@ARTICLE{Gillespie1975,
       author = {{Gillespie}, Daniel T.},
        title = "{An Exact Method for Numerically Simulating the Stochastic Coalescence Process in a Cloud.}",
      journal = {Journal of the Atmospheric Sciences},
         year = 1975,
        month = oct,
       volume = {32},
       number = {10},
        pages = {1977-1989},
          doi = {10.1175/1520-0469(1975)032<1977:AEMFNS>2.0.CO;2},
       adsurl = {https://ui.adsabs.harvard.edu/abs/1975JAtS...32.1977G}
}

@ARTICLE{Umstatter2019,
       author = {{Umst{\"a}tter}, Philipp and {Gunkelmann}, Nina and {Dullemond}, Cornelis P. and {Urbassek}, Herbert M.},
        title = "{Shedding of dust rims in chondrule collisions in the protoplanetary disk}",
      journal = {\mnras},
         year = 2019,
        month = mar,
       volume = {483},
       number = {4},
        pages = {4938-4948},
          doi = {10.1093/mnras/sty3431},
       adsurl = {https://ui.adsabs.harvard.edu/abs/2019MNRAS.483.4938U}
}

@ARTICLE{Simon2015,
       author = {{Simon}, Steven B. and {Grossman}, Lawrence},
        title = "{Refractory inclusions in the pristine carbonaceous chondrites DOM 08004 and DOM 08006}",
      journal = {\maps},
         year = 2015,
        month = jun,
       volume = {50},
       number = {6},
        pages = {1032-1049},
          doi = {10.1111/maps.12452},
       adsurl = {https://ui.adsabs.harvard.edu/abs/2015M&PS...50.1032S}
}

@ARTICLE{Jacquet2014,
       author = {{Jacquet}, Emmanuel and {Thompson}, Christopher},
        title = "{Chondrule Destruction in Nebular Shocks}",
      journal = {\apj},
         year = 2014,
        month = dec,
       volume = {797},
       number = {1},
          eid = {30},
        pages = {30},
          doi = {10.1088/0004-637X/797/1/30},
archivePrefix = {arXiv},
       eprint = {1410.6015},
 primaryClass = {astro-ph.EP},
       adsurl = {https://ui.adsabs.harvard.edu/abs/2014ApJ...797...30J}
}

@ARTICLE{DeMeo2013,
       author = {{DeMeo}, F.~E. and {Carry}, B.},
        title = "{The taxonomic distribution of asteroids from multi-filter all-sky photometric surveys}",
      journal = {\icarus},
         year = 2013,
        month = sep,
       volume = {226},
       number = {1},
        pages = {723-741},
          doi = {10.1016/j.icarus.2013.06.027},
archivePrefix = {arXiv},
       eprint = {1307.2424},
 primaryClass = {astro-ph.EP},
       adsurl = {https://ui.adsabs.harvard.edu/abs/2013Icar..226..723D}
}

@ARTICLE{Colas2020,
       author = {{Colas}, F. and {Zanda}, B. and {Bouley}, S. and {Jeanne}, S. and {Malgoyre}, A. and {Birlan}, M. and {Blanpain}, C. and {Gattacceca}, J. and {Jorda}, L. and {Lecubin}, J. and {Marmo}, C. and {Rault}, J.~L. and {Vaubaillon}, J. and {Vernazza}, P. and {Yohia}, C. and {Gardiol}, D. and {Nedelcu}, A. and {Poppe}, B. and {Rowe}, J. and {Forcier}, M. and {Koschny}, D. and {Trigo-Rodriguez}, J.~M. and {Lamy}, H. and {Behrend}, R. and {Ferri{\`e}re}, L. and {Barghini}, D. and {Buzzoni}, A. and {Carbognani}, A. and {Di Carlo}, M. and {Di Martino}, M. and {Knapic}, C. and {Londero}, E. and {Pratesi}, G. and {Rasetti}, S. and {Riva}, W. and {Stirpe}, G.~M. and {Valsecchi}, G.~B. and {Volpicelli}, C.~A. and {Zorba}, S. and {Coward}, D. and {Drolshagen}, E. and {Drolshagen}, G. and {Hernandez}, O. and {Jehin}, E. and {Jobin}, M. and {King}, A. and {Nitschelm}, C. and {Ott}, T. and {Sanchez-Lavega}, A. and {Toni}, A. and {Abraham}, P. and {Affaticati}, F. and {Albani}, M. and {Andreis}, A. and {Andrieu}, T. and {Anghel}, S. and {Antaluca}, E. and {Antier}, K. and {App{\'e}r{\'e}}, T. and {Armand}, A. and {Ascione}, G. and {Audureau}, Y. and {Auxepaules}, G. and {Avoscan}, T. and {Baba Aissa}, D. and {Bacci}, P. and {B{\v{a}}descu}, O. and {Baldini}, R. and {Baldo}, R. and {Balestrero}, A. and {Baratoux}, D. and {Barbotin}, E. and {Bardy}, M. and {Basso}, S. and {Bautista}, O. and {Bayle}, L.~D. and {Beck}, P. and {Bellitto}, R. and {Belluso}, R. and {Benna}, C. and {Benammi}, M. and {Beneteau}, E. and {Benkhaldoun}, Z. and {Bergamini}, P. and {Bernardi}, F. and {Bertaina}, M.~E. and {Bessin}, P. and {Betti}, L. and {Bettonvil}, F. and {Bihel}, D. and {Birnbaum}, C. and {Blagoi}, O. and {Blouri}, E. and {Boac{\u{a}}}, I. and {Boat{\v{a}}}, R. and {Bobiet}, B. and {Bonino}, R. and {Boros}, K. and {Bouchet}, E. and {Borgeot}, V. and {Bouchez}, E. and {Boust}, D. and {Boudon}, V. and {Bouman}, T. and {Bourget}, P. and {Brandenburg}, S. and {Bramond}, Ph. and {Braun}, E. and {Bussi}, A. and {Cacault}, P. and {Caillier}, B. and {Calegaro}, A. and {Camargo}, J. and {Caminade}, S. and {Campana}, A.~P.~C. and {Campbell-Burns}, P. and {Canal-Domingo}, R. and {Carell}, O. and {Carreau}, S. and {Cascone}, E. and {Cattaneo}, C. and {Cauhape}, P. and {Cavier}, P. and {Celestin}, S. and {Cellino}, A. and {Champenois}, M. and {Chennaoui Aoudjehane}, H. and {Chevrier}, S. and {Cholvy}, P. and {Chomier}, L. and {Christou}, A. and {Cricchio}, D. and {Coadou}, P. and {Cocaign}, J.~Y. and {Cochard}, F. and {Cointin}, S. and {Colombi}, E. and {Colque Saavedra}, J.~P. and {Corp}, L. and {Costa}, M. and {Costard}, F. and {Cottier}, M. and {Cournoyer}, P. and {Coustal}, E. and {Cremonese}, G. and {Cristea}, O. and {Cuzon}, J.~C. and {D'Agostino}, G. and {Daiffallah}, K. and {D{\v{a}}nescu}, C. and {Dardon}, A. and {Dasse}, T. and {Davadan}, C. and {Debs}, V. and {Defaix}, J.~P. and {Deleflie}, F. and {D'Elia}, M. and {De Luca}, P. and {De Maria}, P. and {Deverch{\`e}re}, P. and {Devillepoix}, H. and {Dias}, A. and {Di Dato}, A. and {Di Luca}, R. and {Dominici}, F.~M. and {Drouard}, A. and {Dumont}, J.~L. and {Dupouy}, P. and {Duvignac}, L. and {Egal}, A. and {Erasmus}, N. and {Esseiva}, N. and {Ebel}, A. and {Eisengarten}, B. and {Federici}, F. and {Feral}, S. and {Ferrant}, G. and {Ferreol}, E. and {Finitzer}, P. and {Foucault}, A. and {Francois}, P. and {Fr{\^\i}ncu}, M. and {Froger}, J.~L. and {Gaborit}, F. and {Gagliarducci}, V. and {Galard}, J. and {Gardavot}, A. and {Garmier}, M. and {Garnung}, M. and {Gautier}, B. and {Gendre}, B. and {Gerard}, D. and {Gerardi}, A. and {Godet}, J.~P. and {Grandchamps}, A. and {Grouiez}, B. and {Groult}, S. and {Guidetti}, D. and {Giuli}, G. and {Hello}, Y.},
        title = "{FRIPON: a worldwide network to track incoming meteoroids}",
      journal = {\aap},
         year = 2020,
        month = dec,
       volume = {644},
          eid = {A53},
        pages = {A53},
          doi = {10.1051/0004-6361/202038649},
archivePrefix = {arXiv},
       eprint = {2012.00616},
 primaryClass = {astro-ph.IM},
       adsurl = {https://ui.adsabs.harvard.edu/abs/2020A&A...644A..53C}
}

@ARTICLE{Yokoyama2023,
       author = {{Yokoyama}, Tetsuya and {Nagashima}, Kazuhide and {Nakai}, Izumi and {Young}, Edward D. and {Abe}, Yoshinari and {Al{\'e}on}, J{\'e}r{\^o}me and {Alexander}, Conel M.~O. {\textquoteright}D. and {Amari}, Sachiko and {Amelin}, Yuri and {Bajo}, Ken-ichi and {Bizzarro}, Martin and {Bouvier}, Audrey and {Carlson}, Richard W. and {Chaussidon}, Marc and {Choi}, Byeon-Gak and {Dauphas}, Nicolas and {Davis}, Andrew M. and {Di Rocco}, Tommaso and {Fujiya}, Wataru and {Fukai}, Ryota and {Gautam}, Ikshu and {Haba}, Makiko K. and {Hibiya}, Yuki and {Hidaka}, Hiroshi and {Homma}, Hisashi and {Hoppe}, Peter and {Huss}, Gary R. and {Ichida}, Kiyohiro and {Iizuka}, Tsuyoshi and {Ireland}, Trevor R. and {Ishikawa}, Akira and {Ito}, Motoo and {Itoh}, Shoichi and {Kawasaki}, Noriyuki and {Kita}, Noriko T. and {Kitajima}, Kouki and {Kleine}, Thorsten and {Komatani}, Shintaro and {Krot}, Alexander N. and {Liu}, Ming-Chang and {Masuda}, Yuki and {McKeegan}, Kevin D. and {Morita}, Mayu and {Motomura}, Kazuko and {Moynier}, Fr{\'e}d{\'e}ric and {Nguyen}, Ann and {Nittler}, Larry and {Onose}, Morihiko and {Pack}, Andreas and {Park}, Changkun and {Piani}, Laurette and {Qin}, Liping and {Russell}, Sara S. and {Sakamoto}, Naoya and {Sch{\"o}nb{\"a}chler}, Maria and {Tafla}, Lauren and {Tang}, Haolan and {Terada}, Kentaro and {Terada}, Yasuko and {Usui}, Tomohiro and {Wada}, Sohei and {Wadhwa}, Meenakshi and {Walker}, Richard J. and {Yamashita}, Katsuyuki and {Yin}, Qing-Zhu and {Yoneda}, Shigekazu and {Yui}, Hiroharu and {Zhang}, Ai-Cheng and {Connolly}, Harold C. and {Lauretta}, Dante S. and {Nakamura}, Tomoki and {Naraoka}, Hiroshi and {Noguchi}, Takaaki and {Okazaki}, Ryuji and {Sakamoto}, Kanako and {Yabuta}, Hikaru and {Abe}, Masanao and {Arakawa}, Masahiko and {Fujii}, Atsushi and {Hayakawa}, Masahiko and {Hirata}, Naoyuki and {Hirata}, Naru and {Honda}, Rie and {Honda}, Chikatoshi and {Hosoda}, Satoshi and {Iijima}, Yu-ichi and {Ikeda}, Hitoshi and {Ishiguro}, Masateru and {Ishihara}, Yoshiaki and {Iwata}, Takahiro and {Kawahara}, Kosuke and {Kikuchi}, Shota and {Kitazato}, Kohei and {Matsumoto}, Koji and {Matsuoka}, Moe and {Michikami}, Tatsuhiro and {Mimasu}, Yuya and {Miura}, Akira and {Morota}, Tomokatsu and {Nakazawa}, Satoru and {Namiki}, Noriyuki and {Noda}, Hirotomo and {Noguchi}, Rina and {Ogawa}, Naoko and {Ogawa}, Kazunori and {Okada}, Tatsuaki and {Okamoto}, Chisato and {Ono}, Go and {Ozaki}, Masanobu and {Saiki}, Takanao and {Sakatani}, Naoya and {Sawada}, Hirotaka and {Senshu}, Hiroki and {Shimaki}, Yuri and {Shirai}, Kei and {Sugita}, Seiji and {Takei}, Yuto and {Takeuchi}, Hiroshi and {Tanaka}, Satoshi and {Tatsumi}, Eri and {Terui}, Fuyuto and {Tsuda}, Yuichi and {Tsukizaki}, Ryudo and {Wada}, Koji and {Watanabe}, Sei-ichiro and {Yamada}, Manabu and {Yamada}, Tetsuya and {Yamamoto}, Yukio and {Yano}, Hajime and {Yokota}, Yasuhiro and {Yoshihara}, Keisuke and {Yoshikawa}, Makoto and {Yoshikawa}, Kent and {Furuya}, Shizuho and {Hatakeda}, Kentaro and {Hayashi}, Tasuku and {Hitomi}, Yuya and {Kumagai}, Kazuya and {Miyazaki}, Akiko and {Nakato}, Aiko and {Nishimura}, Masahiro and {Soejima}, Hiromichi and {Suzuki}, Ayako and {Yada}, Toru and {Yamamoto}, Daiki and {Yogata}, Kasumi and {Yoshitake}, Miwa and {Tachibana}, Shogo and {Yurimoto}, Hisayoshi},
        title = "{Samples returned from the asteroid Ryugu are similar to Ivuna-type carbonaceous meteorites}",
      journal = {Science},
         year = 2023,
        month = mar,
       volume = {379},
       number = {6634},
          eid = {abn7850},
        pages = {abn7850},
          doi = {10.1126/science.abn7850},
       adsurl = {https://ui.adsabs.harvard.edu/abs/2023Sci...379.7850Y}
}

@ARTICLE{Barnes2025,
       author = {{Barnes}, J.~J. and {Nguyen}, A.~N. and {Abernethy}, F.~A.~J. and {Bajo}, K. and {Bekaert}, D.~V. and {Bloch}, E. and {Brennecka}, G.~A. and {Busemann}, H. and {Cowpe}, J.~S. and {Crowther}, S.~A. and {Ek}, M. and {Fawcett}, L.~J. and {Fehr}, M.~A. and {Franchi}, I.~A. and {F{\"u}ri}, E. and {Gilmour}, J.~D. and {Grady}, M.~M. and {Greenwood}, R.~C. and {Haenecour}, P. and {Kawasaki}, N. and {Koefoed}, P. and {Krietsch}, D. and {Le}, L. and {Liszewska}, K.~M. and {Maden}, C. and {Malley}, J. and {Marrocchi}, Y. and {Marty}, B. and {Meyer}, L.~A.~E. and {Peretyazhko}, T.~S. and {Piani}, L. and {Render}, J. and {Russell}, S.~S. and {R{\"u}fenacht}, M. and {Sakamoto}, N. and {Sch{\"o}nb{\"a}chler}, M. and {Shollenberger}, Q.~R. and {Smith}, L. and {Thomas-Keprta}, K. and {Verchovsky}, A.~B. and {Villeneuve}, J. and {Wang}, K. and {Welten}, K.~C. and {Wimpenny}, J. and {Worsham}, E.~A. and {Yurimoto}, H. and {Zimmermann}, L. and {Zhao}, X. and {Alexander}, C.~M. O'D. and {Amini}, M. and {Baczynski}, A. and {Bland}, P. and {Borg}, L.~E. and {Burgess}, R. and {Caffee}, M.~W. and {Chaves}, L.~C. and {Clay}, P.~L. and {Dworkin}, J.~P. and {Foustoukos}, D.~I. and {Glavin}, D.~P. and {Hamilton}, V.~E. and {Hill}, D. and {House}, C.~H. and {Huss}, G.~R. and {Ireland}, T. and {Jilly}, C.~E. and {Jourdan}, F. and {Keller}, L.~P. and {Kruijer}, T.~S. and {Lai}, V. and {McCoy}, T.~J. and {Nagashima}, K. and {Nishiizumi}, K. and {Ogliore}, R. and {Ong}, I.~J. and {Reddy}, S.~M. and {Rickard}, W.~D.~A. and {Sandford}, S. and {Saxey}, D.~W. and {Timms}, N. and {Weis}, D. and {Wilbur}, Z.~E. and {Zega}, T.~J. and {DellaGiustina}, D.~N. and {Wolner}, C.~W.~V. and {Connolly}, H.~C. and {Lauretta}, D.~S.},
        title = "{The variety and origin of materials accreted by Bennu's parent asteroid}",
      journal = {Nature Astronomy},
         year = 2025,
        month = dec,
       volume = {9},
        pages = {1785-1802},
          doi = {10.1038/s41550-025-02631-6},
       adsurl = {https://ui.adsabs.harvard.edu/abs/2025NatAs...9.1785B}
}

@ARTICLE{Lauretta2024,
       author = {{Lauretta}, Dante S. and {Connolly}, Harold C. and {Aebersold}, Joseph E. and {Alexander}, Conel M. O'D. and {Ballouz}, Ronald-L. and {Barnes}, Jessica J. and {Bates}, Helena C. and {Bennett}, Carina A. and {Blanche}, Laurinne and {Blumenfeld}, Erika H. and {Clemett}, Simon J. and {Cody}, George D. and {DellaGiustina}, Daniella N. and {Dworkin}, Jason P. and {Eckley}, Scott A. and {Foustoukos}, Dionysis I. and {Franchi}, Ian A. and {Glavin}, Daniel P. and {Greenwood}, Richard C. and {Haenecour}, Pierre and {Hamilton}, Victoria E. and {Hill}, Dolores H. and {Hiroi}, Takahiro and {Ishimaru}, Kana and {Jourdan}, Fred and {Kaplan}, Hannah H. and {Keller}, Lindsay P. and {King}, Ashley J. and {Koefoed}, Piers and {Kontogiannis}, Melissa K. and {Le}, Loan and {Macke}, Robert J. and {McCoy}, Timothy J. and {Milliken}, Ralph E. and {Najorka}, Jens and {Nguyen}, Ann N. and {Pajola}, Maurizio and {Polit}, Anjani T. and {Righter}, Kevin and {Roper}, Heather L. and {Russell}, Sara S. and {Ryan}, Andrew J. and {Sandford}, Scott A. and {Schofield}, Paul F. and {Schultz}, Cody D. and {Seifert}, Laura B. and {Tachibana}, Shogo and {Thomas-Keprta}, Kathie L. and {Thompson}, Michelle S. and {Tu}, Valerie and {Tusberti}, Filippo and {Wang}, Kun and {Zega}, Thomas J. and {Wolner}, C.~W.~V.},
        title = "{Asteroid (101955) Bennu in the laboratory: Properties of the sample collected by OSIRIS-REx}",
      journal = {\maps},
         year = 2024,
        month = sep,
       volume = {59},
       number = {9},
        pages = {2453-2486},
          doi = {10.1111/maps.14227},
archivePrefix = {arXiv},
       eprint = {2404.12536},
 primaryClass = {astro-ph.EP},
       adsurl = {https://ui.adsabs.harvard.edu/abs/2024M&PS...59.2453L}
}

@ARTICLE{Brasser2020,
       author = {{Brasser}, R. and {Mojzsis}, S.~J.},
        title = "{The partitioning of the inner and outer Solar System by a structured protoplanetary disk}",
      journal = {Nature Astronomy},
         year = 2020,
        month = jan,
       volume = {4},
        pages = {492-499},
          doi = {10.1038/s41550-019-0978-6},
       adsurl = {https://ui.adsabs.harvard.edu/abs/2020NatAs...4..492B}
}

@ARTICLE{Morbidelli2024,
       author = {{Morbidelli}, Alessandro and {Marrocchi}, Yves and {Ahmad}, Adnan Ali and {Bhandare}, Asmita and {Charnoz}, S{\'e}bastien and {Commer{\c{c}}on}, Beno{\^\i}t and {Dullemond}, Cornelis P. and {Guillot}, Tristan and {Hennebelle}, Patrick and {Lee}, Yueh-Ning and {Lovascio}, Francesco and {Marschall}, Raphael and {Marty}, Bernard and {Maury}, Ana{\"e}lle and {Tamami}, Okamoto},
        title = "{Formation and evolution of a protoplanetary disk: Combining observations, simulations, and cosmochemical constraints}",
      journal = {\aap},
         year = 2024,
        month = nov,
       volume = {691},
          eid = {A147},
        pages = {A147},
          doi = {10.1051/0004-6361/202451388},
archivePrefix = {arXiv},
       eprint = {2409.06342},
 primaryClass = {astro-ph.EP},
       adsurl = {https://ui.adsabs.harvard.edu/abs/2024A&A...691A.147M}
}

@ARTICLE{Ciesla2006,
       author = {{Ciesla}, Fred J.},
        title = "{Chondrule collisions in shock waves}",
      journal = {\maps},
         year = 2006,
        month = sep,
       volume = {41},
       number = {9},
        pages = {1347-1359},
          doi = {10.1111/j.1945-5100.2006.tb00526.x},
       adsurl = {https://ui.adsabs.harvard.edu/abs/2006M&PS...41.1347C}
}

@ARTICLE{Arakawa2019,
       author = {{Arakawa}, Sota and {Nakamoto}, Taishi},
        title = "{Compound Chondrule Formation in Optically Thin Shock Waves}",
      journal = {\apj},
         year = 2019,
        month = jun,
       volume = {877},
       number = {2},
          eid = {84},
        pages = {84},
          doi = {10.3847/1538-4357/ab1b3e},
archivePrefix = {arXiv},
       eprint = {1904.09580},
 primaryClass = {astro-ph.EP},
       adsurl = {https://ui.adsabs.harvard.edu/abs/2019ApJ...877...84A}
}

@ARTICLE{Krot2009,
       author = {{Krot}, A.~N. and {Amelin}, Y. and {Bland}, P. and {Ciesla}, F.~J. and {Connelly}, J. and {Davis}, A.~M. and {Huss}, G.~R. and {Hutcheon}, I.~D. and {Makide}, K. and {Nagashima}, K. and {Nyquist}, L.~E. and {Russell}, S.~S. and {Scott}, E.~R.~D. and {Thrane}, K. and {Yurimoto}, H. and {Yin}, Q.-Z.},
        title = "{Origin and chronology of chondritic components: A review}",
      journal = {\gca},
         year = 2009,
        month = sep,
       volume = {73},
       number = {17},
        pages = {4963-4997},
          doi = {10.1016/j.gca.2008.09.039},
       adsurl = {https://ui.adsabs.harvard.edu/abs/2009GeCoA..73.4963K}
}

@ARTICLE{Pitjeva2018,
       author = {{Pitjeva}, E.~V. and {Pitjev}, N.~P.},
        title = "{Mass of the Kuiper belt}",
      journal = {Celestial Mechanics and Dynamical Astronomy},
         year = 2018,
        month = sep,
       volume = {130},
       number = {9},
          eid = {57},
        pages = {57},
          doi = {10.1007/s10569-018-9853-5},
archivePrefix = {arXiv},
       eprint = {1810.09771},
 primaryClass = {astro-ph.EP},
       adsurl = {https://ui.adsabs.harvard.edu/abs/2018CeMDA.130...57P}
}

@ARTICLE{Bell1994,
       author = {{Bell}, K.~R. and {Lin}, D.~N.~C.},
        title = "{Using FU Orionis Outbursts to Constrain Self-regulated Protostellar Disk Models}",
      journal = {\apj},
         year = 1994,
        month = jun,
       volume = {427},
        pages = {987},
          doi = {10.1086/174206},
archivePrefix = {arXiv},
       eprint = {astro-ph/9312015},
 primaryClass = {astro-ph},
       adsurl = {https://ui.adsabs.harvard.edu/abs/1994ApJ...427..987B}
}
\bibliographystyle{aasjournalv7}

\clearpage
\appendix
\restartappendixnumbering
\section{Supplementary Tables}

\begin{table*}[h] 
	\centering
	\caption{List of parameters employed in the simulation of gas evolution. Disk parameters not listed in this table are set to the fiducial values of the publicly available version of the code \texttt{DD-Diskevol}}.
	\label{tab:gas} 

	\begin{tabular}{ccc} 
		\\
		\hline
		\multicolumn{1}{c}{Symbol} & \multicolumn{1}{c}{Definition} & \multicolumn{1}{c}{Fiducial Values}   \\ \hline
$M_{\rm{cloud}}$            & Molecular cloud mass             & $1.15\,\mathrm{M_{\odot}}$\\
$\Omega_{\rm{cloud}}$            & Molecular cloud rotation rate             & $10^{-14}\,\mathrm{s^{-1}}$\\
$T_{0}$            & Background temperature             & $10\,\mathrm{K}$\\
$\alpha_{\rm{acc}}$           & Accretion viscosity coefficient                     & $0.005$        \\

$r_{\rm{planet}}$           & Planet location                     & $5\,\mathrm{AU}$        \\
$M_{\rm{planet}}$           & Planet mass                     & $1\,\mathrm{M_{\rm{Jup}}}$        \\
$\alpha_{\rm{acc}}/\alpha_{\rm{peak}}$           & Gap depth                     & $1/1000$        \\ 
$t_{\rm{planet,0}} $          & Planet formation init time                     & $0.6\,\mathrm{Myr}$        \\ 
$t_{\rm{planet,f}} $          & Planet formation final time                     & $1.6\,\mathrm{Myr}$ \\
$t_{\rm{photo}} $          & Photoevaporation activation time                     & $1\,\mathrm{Myr}$        \\

-          & Radial zones                     & $2000$ \\
-          & Standard grid range                      & $40\,\mathrm{R_{\odot}}-10\,\mathrm{pc}$        \\
-          & High resolution grid range                      & $4-7\,\mathrm{AU}$        \\
-          & High resolution steps                     & $0.005\,\mathrm{AU}$        \\
		\hline
	\end{tabular}
\end{table*}

\begin{table*}[h] 
	\centering
	\caption{List of parameters employed in the simulation of dust evolution.}
	\label{tab:dust} 

	\begin{tabular}{ccc} 
		\\
		\hline
\multicolumn{1}{c}{Symbol} & \multicolumn{1}{c}{Definition} & \multicolumn{1}{c}{Fiducial Values}   \\ \hline
$t_{0}$            & Initial time             & $2.19\,\mathrm{Myr}$ or $2\,\mathrm{Myr}$ after CAI  \\
$Z_{0}$            & Vertically integrated initial metallicity             & $5\cdot 10^{-5}$        \\
$v_{\rm{frag}}$            & Fragmentation velocity            & $2\,\mathrm{m\,s^{-1}}$        \\

$\alpha_{\rm{t}}$            & Turbulent parameter            & $10^{-4}$        \\
$\rho_{\rm{fragile}}$            & Internal density of fragile material          & $1.2\,\mathrm{g\,cm^{-3}}$        \\
$\rho_{\rm{rigid}}$            & Internal density of rigid particles            &  $3.3\,\mathrm{g\,cm^{-3}}$        \\
$a_{0}$            & Radii of fragile monomers            &  $10^{-4}\,\mathrm{cm}$        \\

$a^{\rm{min}}_{\rm{rigid}}-a^{\rm{max}}_{\rm{rigid}}$            & Radii range of rigid monomers            &  $10^{-3}-10^{-1}\,\mathrm{cm}$        \\
$\zeta$            & Rigid size power-law index            &  $-3.9$        \\

$\bar{f}_{\rm{ri}}$            & Initial mean rigid mass fraction            &  $0.5$        \\
$\kappa$            & Fragmentation power-law index            &  $1/6$  \\
$\zeta_{\rm{eff}}$            & Planetesimal formation efficiency            &  $10^{-3}$        \\
		\hline
	\end{tabular}
\end{table*}

\clearpage
\section{Supplementary Figures}
\begin{figure}[h]
\centering
\includegraphics[width=0.4\textwidth]{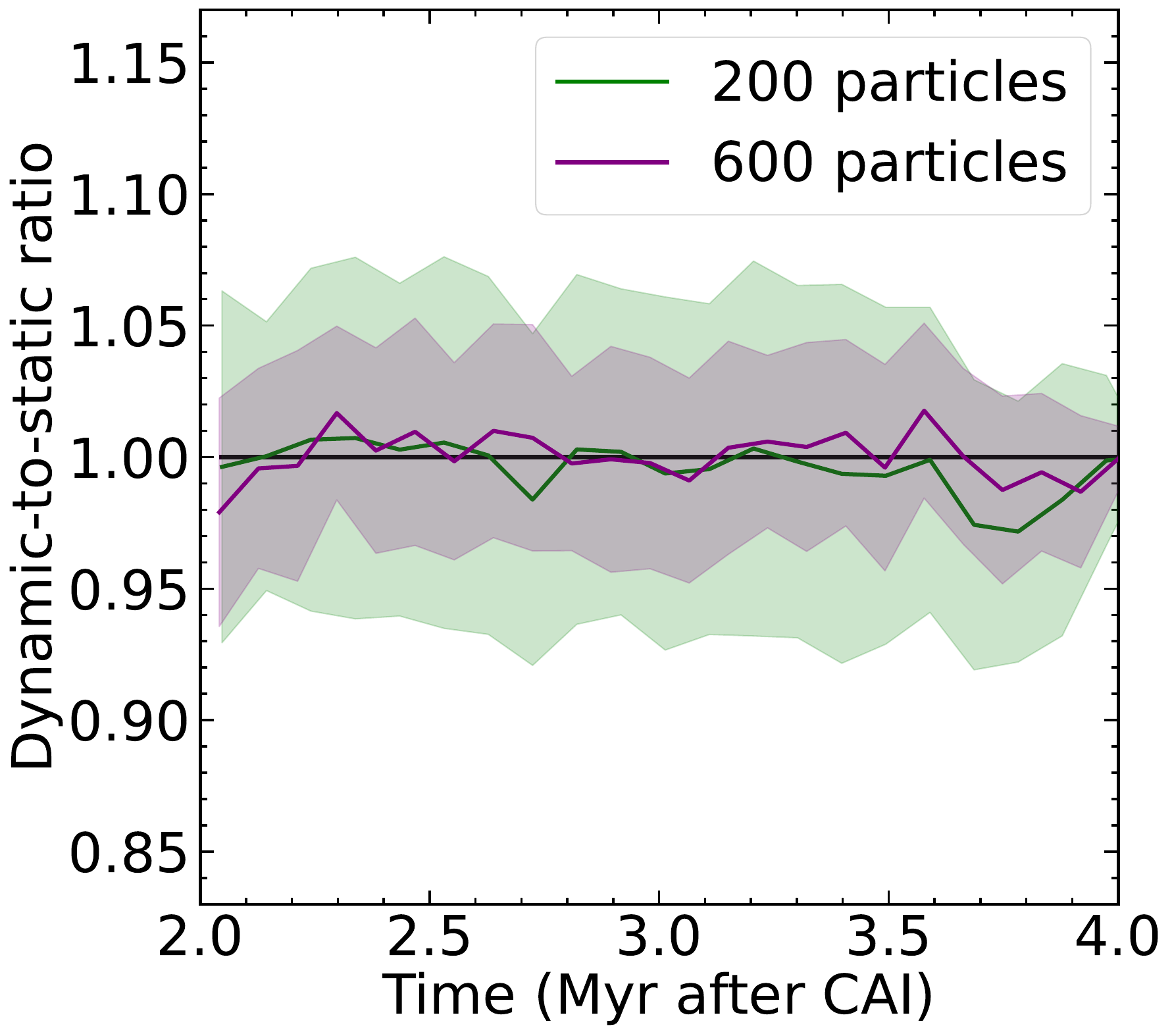}
\caption{Evolution of the total dynamic rigid mass fraction in the zero-dimensional simulation compared to the static value, shown at different resolutions. The solid line indicates the mean ratio calculated over $0.1\,\mathrm{Myr}$ intervals, and the shaded regions show the corresponding standard deviation. In all cases, the dynamic value fluctuates around the static value, with oscillation amplitude decreasing as resolution increases.}\label{fig:mass_conservation}
\end{figure}

\begin{figure}[h]
\centering
\includegraphics[width=0.4\textwidth]{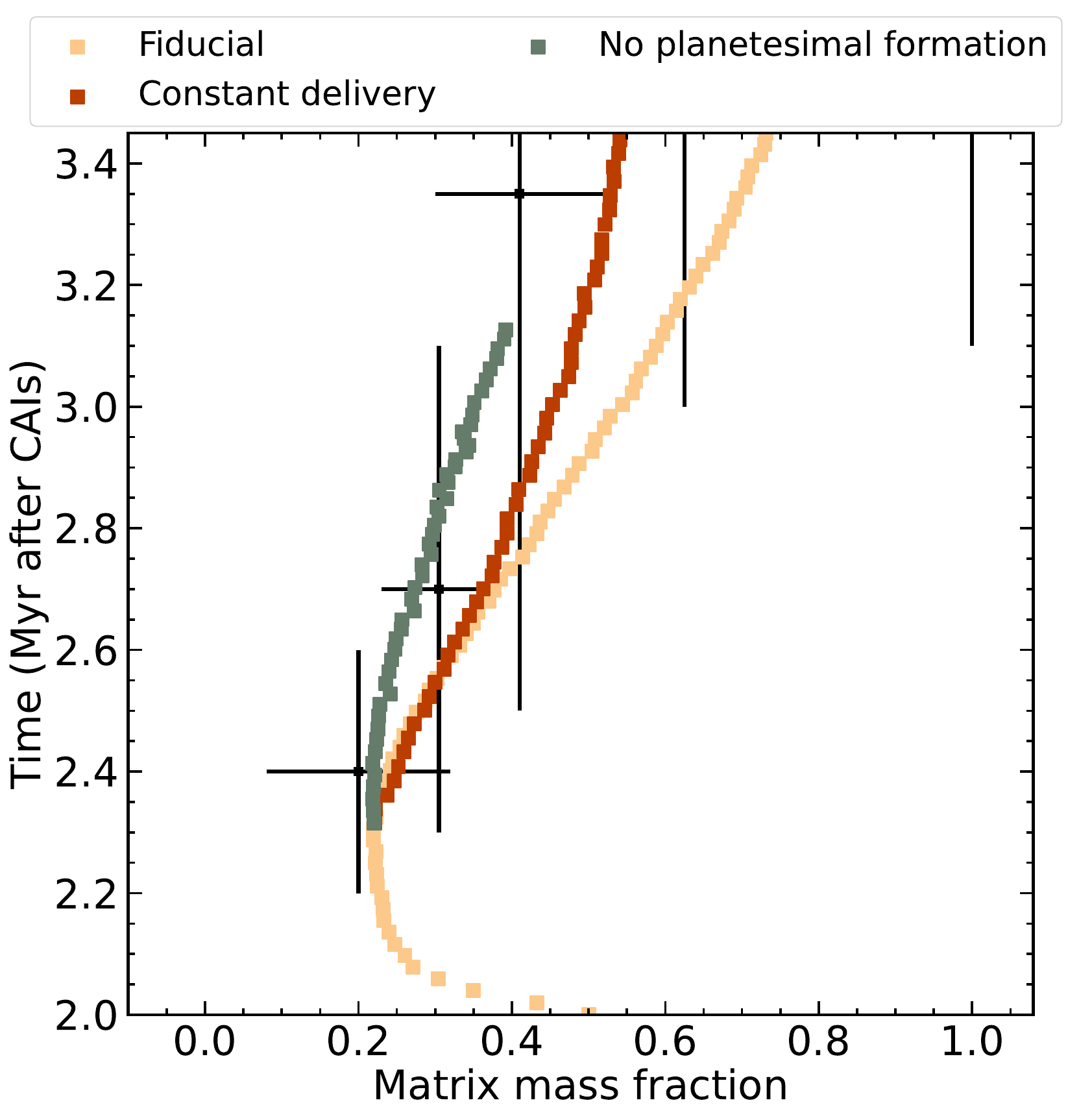}
\caption{Matrix mass fraction of pebbles for three scenarios: (1) the fiducial simulation from the main text; (2) constant rigid fraction of approximately $0.3$ in the feeding flux from $2.3\,\mathrm{Myr}$ onward; and (3) same as (1) but without planetesimal formation. The simulation (3) is presented up to $3.1\,\mathrm{Myr}$ due to the increased computational cost at high dust-to-gas ratios. In (2), the matrix fraction increases less than in the fiducial case, as the replenishing material is richer in rigid particles. In (3), without planetesimal formation, the matrix mass fraction increases more slowly, mainly because the greater dust mass in the trap requires more time to alter its composition at a given feeding rate. These results highlight the importance of including a decreasing rigid particle delivery and planetesimal formation in the model.}\label{fig:differente} 
\end{figure} 

\begin{figure}[h]
\centering
\includegraphics[width=0.4\textwidth]{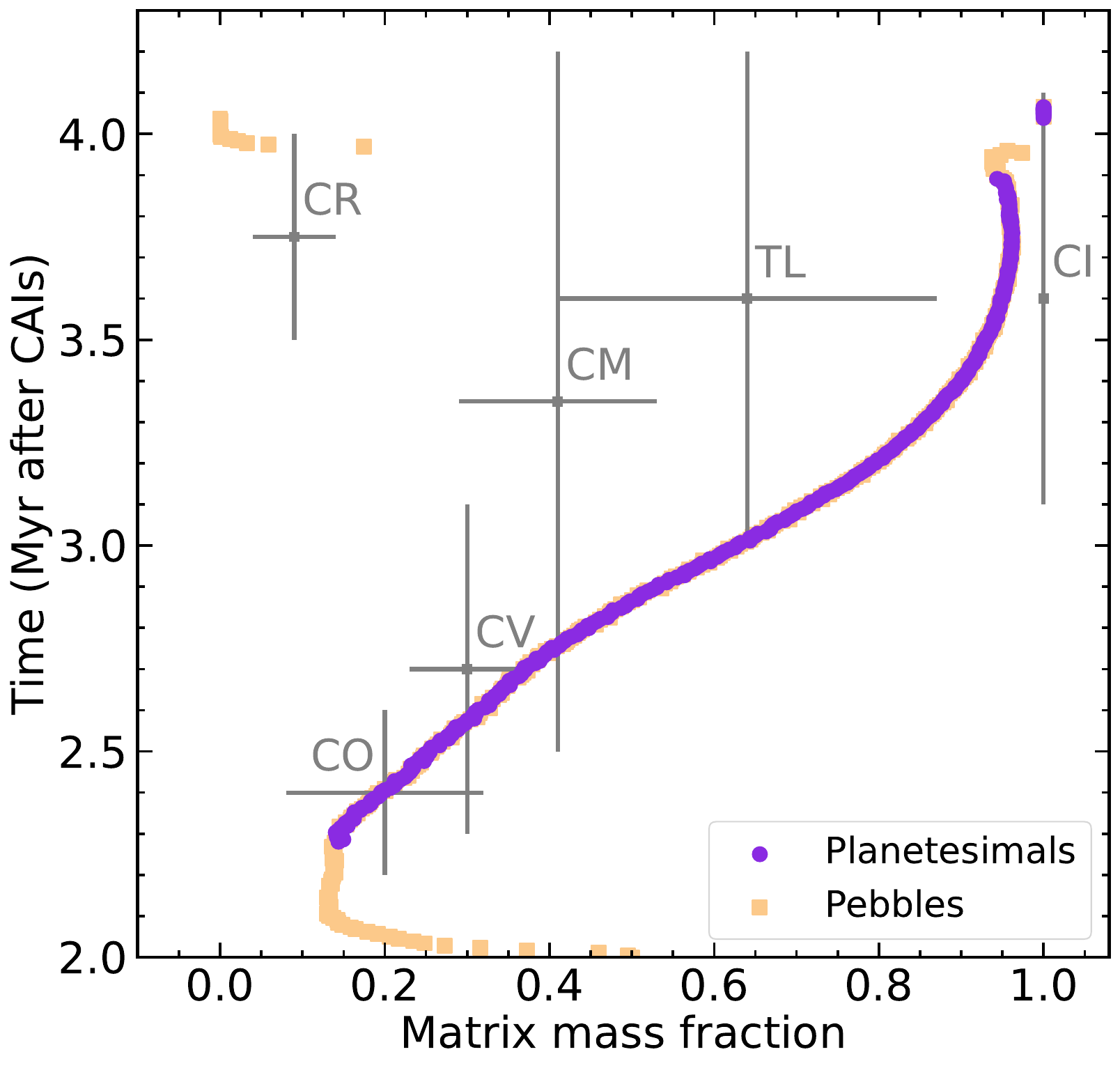}
\caption{Matrix mass fraction of planetesimals for a top-heavy distribution of rigid particle sizes. The initial power-law index is assumed to be $-2.9$, in contrast to the index of $-3.9$ used in Figure~\ref{fig:plt}. Initially, planetesimals have a lower matrix content due to the faster delivery and more efficient trapping of larger rigid particles. However, the matrix fraction increases more steeply over time as the supply of rigid particles to the dust trap becomes depleted. By approximately $3.9-4.\,\mathrm{Myr}$, the mean matrix mass fraction of pebbles in the trap decreases by the same process as in Figures~\ref{fig:CRCI} and \ref{fig:plt}, but the remaining rigid material is insufficient to trigger planetesimal formation.}\label{fig:CR_formation}
\end{figure}

\begin{figure}[h]
\centering
\includegraphics[width=0.4\textwidth]{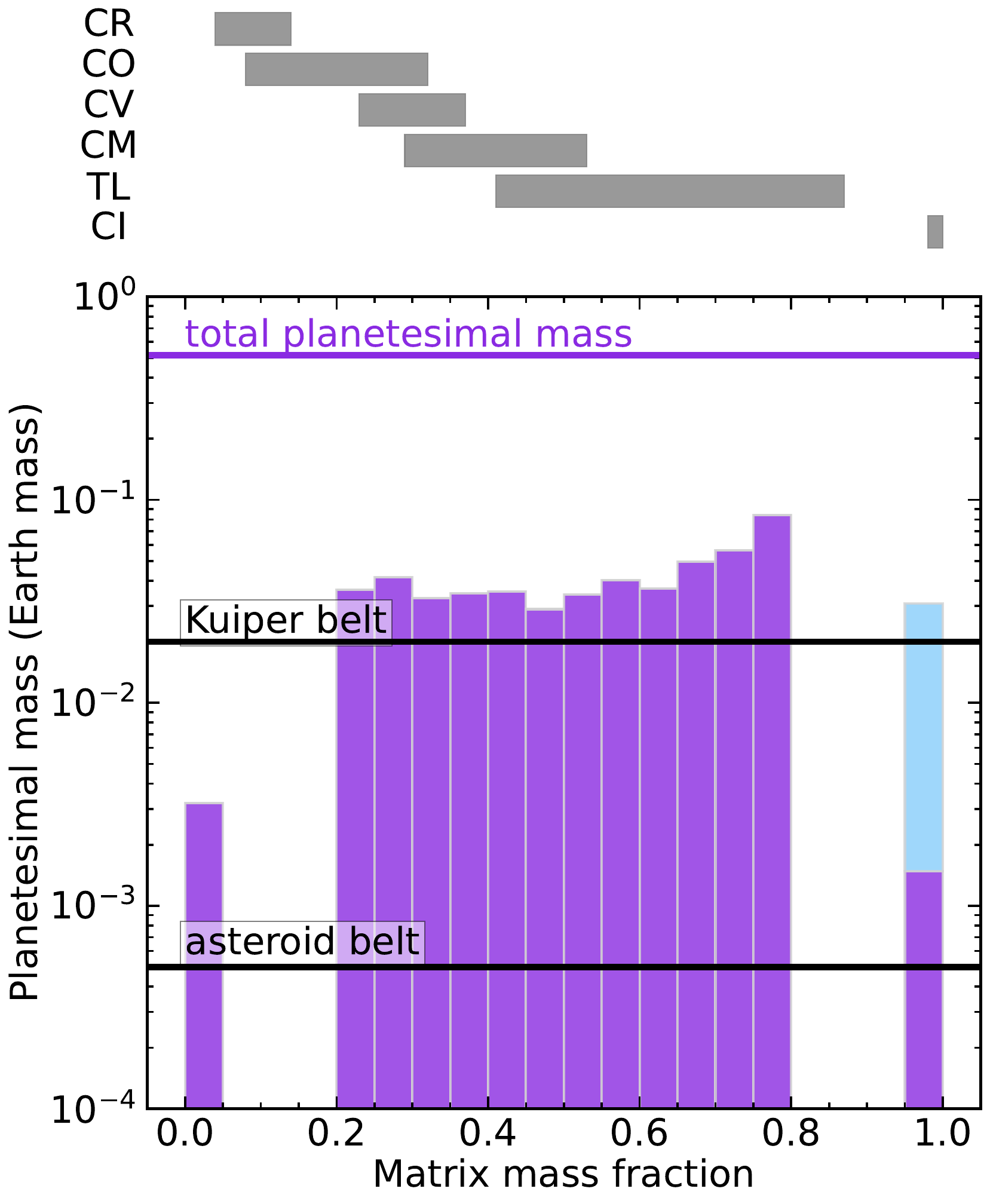}
\caption{Mass of planetesimals formed at the pressure bump as a function of matrix mass fraction in the fiducial simulation (purple bars). The potential planetesimal mass in the outer disk is also estimated by assuming all the remnant material in these regions could form planetesimals (light blue bar). Most planetesimals form with compositions similar to CO, CV, CM, or TL chondrites.  CR-like and CI-like planetesimals at the local simulation account for roughly $0.6\,\%$ and $0.2\,\%$ of the total planetesimal mass. If we account for planetesimals forming in the outer regions as the gap widens, the CI abundance is expected to increase to at most $5\,\%$. The CR abundance may also increase if late-generation chondrules (i.e., conversion of fragile material to rigid one) are included. The simulated total planetesimal mass and the estimated masses of the asteroid belt \citep{DeMeo2013} and the Kuiper belt \citep{Pitjeva2018} are shown for comparison.}\label{fig:planetesimal_mass}
\end{figure}


\clearpage 




\end{document}